\documentclass[useAMS,usenatbib]{mn2e}
\usepackage{graphicx}
\usepackage{amsmath}
\usepackage{amssymb}
\newcommand\nodata{ ~$\cdots$~ }%

\newcommand  \acc     {\ifmmode {\rm km\,s}^{-2} \else km\,s$^{-2}$\fi}

\newcommand  \ergs     {\ifmmode {\rm ergs\,s}^{-1} \else ergs s$^{-1}$\fi}
\newcommand  \ergcms   {\ifmmode {\rm erg~cm}^{-2}\,{\rm s}^{-1}
                        \else erg~cm$^{-2}$\,s$^{-1}$\fi}
\newcommand  \ergcmsA  {\ifmmode{\rm erg\,cm}^{-2}\,{\rm s}^{-1}\,{\rm\AA}^{-1}
                        \else ergs\,cm$^{-2}$\,s$^{-1}$\,\AA$^{-1}$\fi}
\newcommand  \ergcmsHz {\ifmmode{\rm ergs\,cm}^{-2}\,{\rm s}^{-1}\,{\rm Hz}^{-1}
                        \else ergs\,cm$^{-2}$\,s$^{-1}$\,Hz$^{-1}$\fi}
\newcommand  \phcms    {\ifmmode {\rm ph\,cm}^{-2}\,{\rm s}^{-1}
                        \else ph\,cm$^{-2}$\,s$^{-1}$\fi}
\newcommand  \phcmsA   {\ifmmode {\rm ph\,cm}^{-2}\,{\rm s}^{-1}\,{\rm\AA}^{-1}
                        \else ph\,cm$^{-2}$\,s$^{-1}$\,\AA$^{-1}$\fi}

\newcommand\aj{{AJ}}%
\newcommand\apj{{ApJ}}%
\newcommand\apjl{{ApJ}}%
\newcommand\apjs{{ApJS}}%
\newcommand\apss{{Ap\&SS}}%
\newcommand\aap{{A\&A}}%
\newcommand\mnras{{MNRAS}}%
\newcommand\pasp{{PASP}}%
\newcommand\nat{{Nature}}%
\title{
Low-Luminosity 
Active Galactic Nuclei:\\
Are They UV-Faint and Radio Loud? }
\author[Dan Maoz]
{Dan~Maoz$^{1}$\thanks{E-mail: maoz@wise.tau.ac.il 
}\\
$^{1}$School of Physics and Astronomy, 
Tel-Aviv University, Tel-Aviv 69978,
Israel}
\date{\today}

\begin{document}

\maketitle

\label{firstpage}

\begin{abstract}
Low-luminosity active galactic nuclei 
(AGNs) are perceived to be radio loud and
devoid of a ``big blue bump'', 
indicating a transition from a radiatively efficient,
geometrically thin, accretion disc in high-luminosity AGNs, to a 
geometrically thick, radiatively inefficient accretion flow
at low luminosities and accretion rates. 
I revisit the issue of the spectral energy distributions
(SEDs) of low-luminosity AGNs
using recently published, high-angular-resolution
data at radio, ultraviolet (UV), and X-ray wavelengths, for a sample
of 13 nearby galaxies with low-ionization nuclear emission-line
region (LINER) nuclei.
I show that, contrary to common wisdom,
low-luminosity LINERs have significant nonstellar UV flux,
and UV/X-ray luminosity ratios similar, on average, 
to those of Seyfert 1 nuclei $\sim 10^4$ times more luminous. 
The $\alpha_{\rm ox}$ index that quantifies this ratio is in the 
range between -0.8 to -1.4, and is below the extrapolation
to low luminosities of the relation between $\alpha_{\rm ox}$
and UV luminosity observed at higher luminosities.   
In terms of radio loudness, most of the LINERs are indeed 
 radio loud 
(or sometimes even ``super radio loud'') based on
their radio/UV luminosity ratios, when compared
to the most luminous quasars. 
However, the entire distribution of radio loudness 
has been shown to shift to higher radio/UV ratios at low AGN
luminosities. In the context of this global shift, 
some LINERs (the majority) can be considered radio quiet,
and some (from among those with black hole masses $\ga 10^{8.5} M_{\odot}$) 
are radio loud. The SEDs of these low-luminosity ($\sim 10^{40}$\ergs)
AGNs are thus quite similar to those of Seyferts up to 
luminosities of $\sim 10^{44}$\ergs, and there
is no evidence for a sharp change in the SED 
at the lowest luminosities.
Thin AGN accretion discs may therefore persist at low accretion rates, 
in analogy to some recent findings for Galactic stellar-mass accreting 
black holes.
\end{abstract}


\section{Introduction}

Supermassive black holes (BHs), whose existence has been postulated 
since the discovery of quasars, are now believed to reside
in the nuclei of all massive galaxies, with a correlation 
between the masses of the BHs and the masses of their host bulges (as
evidenced by the stellar luminosities or the velocity dispersions of the
bulges; e.g., H\"aring \& Rix 2004; Tremaine et al. 2002).
Following an argument by Soltan (1982), 
the active galactic nucleus (AGN) luminosity density
from recent X-ray and optical surveys, integrated over cosmic time,
when compared to the present-day space density of nuclear BHs,
provides strong evidence that the BHs have grown primarily 
during relatively brief ($\sim 10^8$~yr) 
active periods when they accreted in a radiatively 
efficient mode as Seyfert nuclei and quasars (Marconi et al. 2004; 
Shankar et al. 2004, Cao 2007).
The favored mechanism for achieving the required high 
rest-mass-to-energy conversion efficiencies ($\epsilon \approx 0.1$)
is a geometrically thin, optically thick, accretion disc
(Shakura \& Sunyaev 1973). A less efficient accretion mode 
during the Seyfert/quasar phase would
overpredict the currently observed BH space density.
Semi-direct evidence for the existence of such discs in AGNs has come
from the presence of an optical to ultraviolet (UV) 
``big blue blump'' in the spectral
energy distributions (SEDs) of luminous AGNs (e.g., Shang et al. 2005;
Kishimoto et al. 2005) -- presumably the thermal 
radiation from the disc (Shields 1978; Malkan 1983), 
as well as from the profiles of iron X-ray emission lines,
interpreted as gravitationally redshifted and relativistically 
Doppler-broadened and -boosted emission from the inner regions of the
disc (see Nandra et al. 2006, for a recent review).  
Estimates of the BH masses in Seyferts and quasars indicate,
furthermore, that the thin discs are often radiating at 
luminosities near the Eddington limit, $L_E$, corresponding to the
BH mass (Kaspi et al. 2000; Kollmeier et al. 2006; Netzer \&
Trakhtenbrot 2007). 

Compared to the radiatively efficient, high-rate,
accretion mode of BHs in their short-lived, high-luminosity, phases, 
much less is known about BH behavior during their long, quiescent, 
stages. In the presence of a central BH, and given the ubiquity of
gas from the mass loss accompanying normal stellar evolution of the bulge
population, it appears unavoidable that accretion on to the BH at some
finite rate still takes place (e.g., Soria et al. 2006a,b). However,
it has been a challenge to explain how this accretion can be
accompanied by only the feeble radiative signatures that are observed
in non-active galaxies. 
During their quiescent stage, BHs may switch
to a different accretion mode, characterized by a low accretion rate
and low-radiative efficiency. In most models of such radiatively inefficient
accretion flows (RIAFs; see, e.g., Yuan 2007, and references therein), 
the kinetic energy associated with the gas
is either advected with the matter into the BH, or redirected into 
an outflow. The funnels in the 
geometrically thick structures, such as tori, invoked
by such models, also provide a plausible mechanism for collimating
axial outflows and jets. Observationally, AGNs show
some evidence for a transition from
a low-accretion-rate state, with an associated low luminosity and a hard X-ray
spectrum, to a high-accretion-rate, high luminosity, soft-X-ray state,
analogous to behavior observed in Galactic stellar-mass accreting BHs
(e.g., Done \& Gierli{\'n}ski 2005).
However, direct observational evidence for
the existence of RIAFs has been sparse, and some authors have argued
against such structures based on the observed spectral slopes
in specific wavelength regimes (e.g., Done \& Gierli{\'n}ski 2005; 
K\"ording, Falcke, \& Corbel 2006), when
 compared to the predictions of RIAF models.

Understanding the quiescent BH stage is important for a 
variety of reasons. It is not known what triggers the 
active stage of luminous AGNs, with galaxy interactions, or
tidal disruption and accretion of stars being 
favorite contenders (e.g., Volonteri et al. 2006), 
and  details about the quiescent stage could
provide clues. The existence of the BH-bulge correlations and results
of galaxy-formation simulations have led to proposals that AGNs provide
``feedback'' that can regulate star formation and can affect
galactic structure (e.g., Springel et al. 2005; Silk 2005). 
These ideas have been reinforced by the recent
discovery of large cavities of hot gas that have been formed 
in the intracluster medium by jets from the nuclei of 
central cluster galaxies (e.g., Dunn \& Fabian 2006; Nusser et
al. 2006). This feedback could well take place mainly
during the quiescent stage of BH activity (e.g., Heinz et al. 2007).

The low-level nuclear activity of the quiescent stage
can manifest itself observationally 
in the form of compact,
often-variable, sources of radio, UV, or X-ray emission, 
or as emission-line nuclei with line
ratios uncharacteristic of stellar excitation, or with broad Balmer
lines reminiscent of Seyfert 1 spectra (see below). 
Jets, generally seen in radio,
but sometimes in other bands as well  (e.g., as in M87) are another
unambiguous indicator of nuclear activity seen in some
galaxies. The weakness of these radiative signals from quiescent
BHs make their study difficult, almost by definition. Problems
include confusion with brighter, non-AGN, components, obscuration by dust, 
and selection effects.
Fortunately, it is becoming clear that a large fraction of all galactic nuclei
do show some weak signs of activity that can be associated with the central 
BH, and in this sense most galaxies can be considered to 
host a low-luminosity AGN.
The ubiquity and nearness of these objects (including the one in our
own Galaxy) often permits separating their weak signals from the various
possible backgrounds, such as radiation from stars in the optical to the 
UV range, and from circumnuclear accreting binaries in X-rays.
In radio, high-angular resolution can help separate between jet 
emission and an unresolved nuclear source that could be coming 
directly from the accretion flow.  

Probably the most common manifestation of low-luminosity 
AGN appears in the form
low-ionization
nuclear emission-line regions (LINERs; Heckman 1980), 
which are detected in the
nuclei of a large fraction of bright nearby galaxies
(Ho et al. 1997; Kauffmann et al. 2003).
As their name implies, LINERs are characterized by 
collisionally excited lines of neutral and singly-ionized species,
indicators of the presence of hot but largely neutral gas.
Theoretical studies that attempted to model LINER spectra 
have generally agreed that LINERs are photoionized objects.
The ionizing sources could be either an AGN-like UV-to-X-ray source,
or a cluster of massive stars of the right mix to 
produce a hard-enough  spectrum (e.g., Barth \& Shields 2000; Kewley et
al. 2006). 
Although the nature of LINERs and their relation, if any, to AGNs has been
debated for several decades, it is now becoming clear that at least a large
fraction of LINERs\footnote{I will use the term 
LINERs here
to denote only the population of compact nuclei with such emission,
and will ignore a 
population of ``extended LINERs'' in which a LINER spectrum
is produced under apparently different circumstances. See Sturm et al. (2006)
for some recent characterizations of extended LINERs, which 
are relatively luminous in IR bands, and distinct in several ways from
the compact sources under discussion here.} are indeed AGNs (see,
e.g., discussion in Ho et al. 2003). Recently,
Kewley et al. (2006) have used Sloan Digital Sky Survey (SDSS) data to show 
that LINERs occupy a well-defined ``cloud'' in emission-line ratio
diagnostic diagrams, a cloud that is distinct from the regions 
occupied by H~II and Seyfert nuclei. These results largely overcome
previous sentiments that LINERs are a ``mixed bag'' of objects,
with some photoionized by stars, some by AGNs, and some perhaps
excited by shocks. By assuming a BH mass for every galaxy based
on the BH-bulge mass relation, Kewley et al. (2006) further showed that 
the transition from the Seyfert region to the LINER region in 
the diagnostic diagrams corresponds to a decrease in the Eddington
ratio, $L/L_E$, confirming earlier conclusions by Ho (2004).
If this decrease is accompanied by a hardening of the
ionizing spectrum, the spectral transition can also be explained.    

In terms of actually detecting the ionizing continuum source in
LINERs, observations at UV wavelengths, where the bright background
from the bulge stellar population basically disappears, have 
proved to be useful. 
{\it Hubble Space Telescope} (HST) 
imaging showed that some 25 per cent of LINERs have compact, 
generally unresolved 
(i.e., $\la$ few pc), bright UV sources in their nuclei 
(Maoz et al. 1995; Barth et al. 1998).
Optical HST imaging of 14 LINERs (Pogge et al. 2000) revealed that
in every case where a compact UV nucleus had not been detected,
obscuration of the nucleus by circumnuclear dust was apparent.  
This strongly suggests that most nearby LINERs 
(including the 75 per cent that are ``UV-dark'', i.e., those that
do not reveal a nuclear UV source at HST sensitivity, $\sim
10^{-17}~\ergcmsA$) 
likely have such a nuclear UV source. 

Maoz et al. (2005; M05) 
monitored with the HST {\it Advanced Camera for Surveys} (ACS)
a sample of 17 compact UV LINER
nuclei at 2500~\AA~ and 3300~\AA. 
They found that all but three of the objects varied
in UV brightness on month-long time-scales, correlated between both UV
bands, or on decade-long time-scales (by comparison to
flux levels measured previously [1993--2000] for these objects at 
bandpasses similar to the ACS 2500~\AA~ band, usually at 2300~\AA), or both.
Months-scale variation amplitudes were typically $\sim 10$ per cent,
while decadal variations were by a factor of a few.
This result argues for a nonstellar UV source and,
by extension to the ionizing far-UV range, for AGN excitation
of the LINER emission lines.  
The variable UV flux provides a lower limit on the nonstellar 
contribution to the UV luminosity of each object.  

In the radio,
at Very Large Array (VLA) resolution ($0\farcs 1\approx $ tens of pc), 
about half of LINERs display
unresolved radio cores at 2~cm and 3.6~cm
 (Nagar et al. 2000, 2002). With 
Very Long Baseline Interferometer
(VLBI) resolution at 6~cm ($\sim 1$ pc),
these cores remain unresolved, strongly arguing for the presence of an AGN
(Falcke \& Biermann 1999; Falcke et al. 2000). 
The radio core fluxes have been found
to be variable by factors of up to a few 
in about half of the $\sim 10$ LINERs observed multiple times
over 3 years (Nagar et al. 2002). 

At X-ray energies, {\it Rosat} HRI images showed compact ($<5''$) soft X-ray
emission in 70 per cent of LINERs and Seyferts (Roberts \& Warwick 2000) which, 
when observed with {\it ASCA}, were found to have a nonthermal
2--10 kev spectrum (e.g., Terashima et al. 2000). Arcsecond-resolution
{\it Chandra} 
observations by Terashima \& Wilson (2003) of 11 LINERs, each of which 
was
preselected to have a radio core, revealed an X-ray nucleus in all but
one case, and the nuclei  were generally (but not always) unresolved.
Most recently, Flohic et al. (2006) found an X-ray nucleus in 12 out 
of 19 LINERs observed with {\it Chandra}. They argued that, in most cases,
the observed X-ray AGN is not luminous enough to power the H$\alpha$ 
emission through photoionization. However, 
this conclusion depends on the assumed extrapolation of the
0.5--10 keV spectrum to far-UV energies. In fact, Flohic et al. showed
that the LINERs are consistent with the best-fitting relation found
by Ho et al. (2001) between 2--10~keV and H$\alpha$ luminosities
for quasars and Seyfert 1 galaxies. 
This implies that, if LINERs and luminous AGNs have similar
UV-to-X-ray SEDs, then there is no photoionization energy budget 
problem in LINERs.

The identification of the nonstellar emission components in LINERs
permits obtaining a picture of the 
nuclear SEDs of these low-luminosity AGNs across the electromagnetic spectrum.
Comparison of the SEDs to theoretical predictions for different accretion
modes may be the most promising avenue for understanding how
BHs ``sleep''. 

Ho (1999) compiled and studied SEDs for seven low-luminosity AGNs,
including four LINERs, a Seyfert, and two borderline LINER-Seyfert
cases (in the new classification scheme of Kewley et al. [2006],
both borderline cases, M81 and NGC~4579, are unambiguous LINERs).
His main conclusion was that low-luminosity AGN SEDs are markedly 
different from those of luminous AGNs, in that underluminous objects 
have a weak or absent big blue bump, 
and are ``radio loud'' in terms of the ratio 
of luminosity in the radio relative to other bands. 
However, several problems could cast doubt on these conclusions.
In two of the objects (NGC~4261 and NGC~4374) the nucleus is obscured 
in the UV by conspicuous patches of circumnuclear dust, resulting
in non-detections in this band. In a third object, NGC~6251, there are no
space-UV data. Thus, 
four objects (M81, NGC~4594, NGC~4579, and M87) 
with useful measurements remain 
in the critical space-UV region. A study by Ho et al. (2002) included
three additional LINERs (NGC~1097, NGC~4203, and NGC~4450), plus M81, and
reached similar conclusions. Both of these studies
included HST optical-band measurements in the SEDs, and considered
optical-to-UV spectral slopes. I will
argue (\S\ref{comparisonsection})
that this is risky because, even at HST resolutions, 
nuclear starlight from the centrally peaked, often cusped, 
bulge light distributions 
can contaminate the optical measurements. This, and/or small
amounts of reddening in the UV, can distort the SED in optical bands. 
  
Ho (2002) studied further the radio-loudness issue, and showed
that radio loudness anticorrelates strongly with Eddington ratio.
The ratio of radio to optical luminosity in AGN samples,
 as a function of Eddington ratio, was recently studied also by 
Chiaberge et al. (2005) and Sikora et al. (2007). Both of
these studies were again based on optical measurements of low-luminosity
nuclear fluxes, which are susceptible to contamination by non-AGN
components, even at HST resolution. The sample of Chiaberge et
al. (2005) included 21 LINERs, and that of Sikora et al. (2007)
has four LINER nuclei.

The SEDs of several low-luminosity AGNs have been compiled
and analysed also on an individual basis: 
M81 and NGC~4579 (Quataert et al. 1999); NGC 6166 (Di Matteo et al. 2001);
NGC 4258 (Yuan et al. 2002); IC 4296 and NGC~1399 (Pellegrini et al. 2003a);
M87 (Wilson \& Yang 2002; Di Matteo et al. 2003; Sabra et al. 2003);
NGC 4594 (Pellegrini et al. 2003b);
NGC~3998 (Ptak et al. 2004); and NGC~4565 (Chiaberge et al. 2006). 
Most of these analyses did not include
UV data, and sometimes did include optical and IR data,
which involve the risks mentioned above.
Six of the objects above will be re-analysed in the present work.
     
In this paper, I revisit the SEDs of low-luminosity AGNs, 
motivated by several recent developments. First, a larger sample
of LINERs having accurate HST/ACS
 UV photometry is now available.
Furthermore, the observed variable fraction of the UV flux found by
M05 in these
objects provides a firm lower limit on the nonstellar AGN flux.
Second, high-resolution X-ray measurements with {\it Chandra}
and {\it Newton XMM} exist for most of these objects, permitting better
isolation of the compact central X-ray source. Finally, recent
statistical studies of the spectral properties of AGNs (e.g., Steffen
et al. 2006; Sikora et al. 2007; Panessa et al. 2006,2007)
allow a clearer comparison of SEDs as a function of luminosity
and Eddington ratio,
and thus give a better view of
 low-luminosity AGNs in the greater AGN context.  

\section{Sample}

My objective in this work is to investigate the 
luminosity ratios between radio, UV, and X-ray emission
in low-luminosity LINER-type AGNs. High-angular-resolution 
UV measurements, possible only with HST, have been carried
out for relatively few galaxies, and in only 25 per cent of
the LINERs among them is the UV nucleus unobscured by
circumnuclear dust (see \S1, above).    
I therefore start with the 17 galaxies
with known central UV nuclei, monitored in 2002--2003 with HST
by M05. Although originally selected
by M05 as LINERs, 
three of the 17 objects in the M05 sample were classified by Ho et
al. (1997) as Seyferts, since their narrow
emission line ratios [OIII]~$\lambda 5007$/H$\beta$ were
above the defining border between LINERs and Seyferts by $\sim 30-40$ per cent
(for M81 and NGC~3486) and by a factor  $\sim 3$ (for NGC~4258).
However, in the new classification scheme proposed by Kewley et al. (2006),
and which separates well LINERs from Seyferts, 
M81 is clearly a LINER. NGC~3486 is a Seyfert according to its
[OIII]/H$\beta$ and [OI]/H$\alpha$ ratios, but is just on the 
LINER-Seyfert boundary in the 
[OIII]/H$\beta$ vs. [SII]/H$\alpha$ diagram, and is thus an ambiguous
case. NGC~4258 is a Seyfert according to its [SII]/H$\alpha$ and 
[OI]/H$\alpha$ ratios.
I will therefore consider M81 a LINER, NGC~3486
a borderline LINER/Seyfert, and NGC~4258 a Seyfert nucleus.

From this initial selection of 17 objects, I exclude from the
sample the following galaxies. NGC~4258 is excluded, because, contrary
to all the other objects, it is a Seyfert, rather than a LINER.
Furthermore, the morphology of its nuclear region in HST images 
suggests that the nucleus is partly obscured by dust (Pogge et al. 2000).  
This suspicion is reinforced by the high absorbing column ($N_{\rm H}\sim
 10^{23}~{\rm cm}^{-2}$) fitted by 
Pietsch \& Read (2002) to their XMM-{\it Newton} measurements of this object.
The UV flux measured by M05 therefore likely underestimates by a large
factor the unabsorbed UV flux, and using it could lead to a 
distorted picture of the true SED. Finally, the 
unabsorbed nuclear 2--10~keV flux of 
$12\times 10^{-12}$~\ergcms, found by Pietsch \& Read (2002)
corresponds to a luminosity of $\sim 10^{42}$\ergs, typical of Seyfert
nuclei, but 100 times larger than the typical luminosities
of the rest of the LINER sample.  
NGC~6500 is excluded because it has
no clear nuclear UV source, and the nucleus may be
obscured (see M05); and NGC~4569 and NGC~5055 are excluded because,
 for both of these
cases, there are multiple lines of evidence that a young nuclear star
cluster, rather than an AGN, dominates the energetics: nondetection
in radio (Nagar et al. 2000); 
a resolved UV source (Barth et al. 1998; M05); 
a UV spectrum dominated by clear signatures
of massive stars (Maoz et al. 1998); 
non-variability in the UV on both long and
short time-scales (M05); and a very low X-ray luminosity (Ho et
al. 2001; Flohic et al. 2006).

The remaining 13 objects in the sample
include both LINERs having broad 
H$\alpha$ wings (which I will designate ``LINER 1s''), 
and those having only narrow emission 
lines (``LINER 2s'')\footnote{Ho et
  al. (1997) designated LINERs with broad H$\alpha$ wings as
  LINER~1.9 objects. Since this is the only kind of type-1 LINER in the
  sample of Ho et al. (i.e., there are no known examples of
  LINER~1.2, 1.5, etc.), I will simply refer to LINER~1.9s as
  LINER~1s.}
 The nucleus of NGC~4552, has broad
components in both the permitted and the forbidden lines 
(Cappellari et al.1999), and 
therefore does not fall easily into either the type-1 or type-2
category. All but one of the 13 objects, were found to be 
UV-variable by M05, on short or long time-scales, or both (the
exception is NGC~3486). Both types, 
LINER~1s and LINER~2s, are variable. M05 found a rough trend in UV colour,
in the sense that LINER~2s have a redder flux ratio, 
$f_{\lambda}({\rm 250W})/f_{\lambda}({\rm 330W})$, compared to
LINER~1s, by a factor of $\sim 2$. 
Table~\ref{table1} lists the objects in the sample and summarizes some of 
their properties. 

\begin{table*}
\scriptsize
\begin{minipage}{\textwidth}
\begin{tabular}{l|c|c|c|c|r|r|r|r|r|r|r|r|r|r|r}
\hline
\hline
{Object} &
{L} &
{$D$} &
{$A_B$} &
{$M_{\rm BH}$} &
{$f_{\nu}$} &
{$f_{\nu}$} &
{$f_{\nu}$} &
{$f_{\nu}$} &
{$f_{\lambda}$l.l.} &
{$f_{\lambda}$} &
{$f_{\lambda}$l.l.} &
{$f_{\lambda}$} &
{$f$}&
{$f$}&
{$\Gamma$} \\
{} &
{} &
{Mpc} &
{mag} &
{} &
{6~cm} &
{3.6~cm} &
{2~cm} &
{0.7~cm} &
{3300~\AA} &
{3300~\AA} &
{2500~\AA} &
{2500~\AA} &
{0.5--2~keV} &
{2--10~keV} &
{} \\
{(1)} &
{(2)} &
{(3)} &
{(4)} &
{(5)} &
{(6)} &
{(7)} &
{(8)} &
{(9)} &
{(10)} &
{(11)} &
{(12)} &
{(13)} &
{(14)} &
{(15)} &
{(16)} \\
\hline
NGC~~404&2&  3.0&0.253& 5.3& \nodata&   $<$0.3&   $<$0.4&  $<$3.0& \nodata&  85.0&  $>$41.0&  134.0&   1.5& \nodata& 1.8 \\
NGC~1052&1& 18.0&0.114& 8.1&2410.0& 2390.0&2090.0& \nodata& 670.0&   9.0&   $>$7.3&   15& \nodata& 240.0& 1.4 \\
M81     &1&  3.6&0.346& 7.8& \nodata& 132.0& 165.0& \nodata& $>$14.0& 130.0&  $>$16.0& 200.0& \nodata&1000.0& 1.8 \\
NGC~3368&2& 10.7&0.109& 7.4& \nodata& \nodata&   $<$0.6& \nodata& \nodata&  30.0&  $>$17.0&  22.0& \nodata&  16.0& 1.8 \\
NGC~3486&2&  7.4&0.093& 6.2&   $<$0.1& \nodata& \nodata& \nodata& \nodata&  18.2& \nodata&  10.8& \nodata&   $<$0.5& 1.8 \\
NGC~3642&1& 27.5&0.046& 7.1& \nodata& \nodata&   $<$0.4& \nodata&   $>$1.8&  22.1&   $>$2.0&  24.5&  26.0& \nodata& 1.8 \\
NGC~3998&1& 13.1&0.069& 8.4&  83.0& \nodata& \nodata& \nodata&  $>$28.0& 153.0&  $>$38.0& 199.0& \nodata&1100.0& 1.9 \\
NGC~4203&1& 15.1&0.052& 7.0&   8.1&   8.5&  10.2&   9.9&  $>$13.0&  37.0&  $>$57.0&  58.0& 119.0& 125.0& 1.8 \\
M87      &2& 15.4&0.096& 9.5& \nodata& \nodata&3000.0& \nodata& \nodata&  48.0&  $>$45.0& 100.0&  59.0& \nodata& 2.2 \\
NGC~4552&?& 15.4&0.177& 8.5&  99.5& \nodata&  59.0& \nodata&   $>$0.3&   1.5&   $>$0.4&   2.0&   5.0&   6.0& 1.8 \\
NGC~4579&1& 21.0&0.177& 7.8&  19.2&  16.2&  15.3&  14.6& $>$2.9&  42.0&  $>$77.0& 110.0& 300.0& 380.0& 1.7 \\
NGC~4594&2&  9.1&0.120& 9.0& 123.0& \nodata& 100.0& \nodata&   $>$1.7&  15.3&   $>$5.1&   12.0&  55.0& 130.0& 1.9 \\
NGC~4736&2&  4.9&0.076& 7.1& \nodata&   1.7&   1.7& \nodata& \nodata&  73.0&   $>$2.0&  39.0& \nodata&  27.0& 1.6 \\
\hline
\end{tabular}
\end{minipage}
\caption{Adopted Data. See \S\ref{data} for sources and references
for all data. Column header explanations: (2)- Type-1 or type-2 LINER, depending on presence or
  absence, respectively, of broad H$\alpha$. NGC~4552, which does
  not fall easily into either category (see text), is
  marked with a ``?''; (3) - distance
(4) - $B$-band Galactic extinction; (5) - log of black hole mass,
in Solar-mass units;
 (6)-(9) - monochromatic radio flux per unit frequency, in mJy. upper
limits are $3\sigma$;
 (10)-(13) - monochromatic UV flux per unit wavelength, in 
units of $10^{-17}\ergcmsA$. Columns headed ``l.l.'' give a lower
limit on the nontellar contribution based on the variable flux;
(14)-(15) absorption-corrected X-ray flux, in units of $10^{-14}\ergcms$;
(16)- X-ray photon index. When unknown, $\Gamma=1.8$ is assumed.
}
\label{table1}
\end{table*}

\section{Data Compilation}
\label{data}
In my analysis, I will use only radio, UV, and X-ray data, and will
ignore optical and IR measurements. Although high-resolution
HST optical and
{\it Spitzer} IR measurements are available for several of the LINERs
in the sample, contamination by starlight, or by
circumnuclear dust heated by stars, could distort the SED 
of the central accreting structure (see \S\ref{comparisonsection}). Only 
a measurement of {\it variable} optical and IR flux, that could
therefore be associated unambiguously with an AGN, may provide 
in the future useful data for this kind of analysis. 

I describe below some general characteristics and considerations
for selecting the data used
in every band. I then relate the details of the data sources
and their choices
for each object individually. The data adopted for every galaxy
are summarized in Table~\ref{table1}. I do not cite the formal 
measurement errors reported by the original workers.
In my subsequent analysis of LINER SEDs, these errors
are largely irrelevant, as they are dwarfed by bigger, systematic,
errors: non-simultaneity of the measurements in the presence
of variability (see below); angular resolution effects in radio bands,
leading to the ``resolving out'' of flux at high resolutions (see,
e.g., Best et al. 2005);
uncertain corrections for internal extinction
in the UV; and, in X-rays, uncertain absorption corrections,
uncertainties in multi-component spectral modeling, and
calibration uncertainties.  

\subsection{Radio Data}
Very Large
Baseline Array (VLBA) and VLBI imaging has
 shown that at least some of the radio emission in 
LINERs is contributed by jets, rather than by an actual accretion flow
(e.g., Falcke et al. 2000; Nagar et al. 2002) 
For radio data, I have therefore 
searched the literature for measurements with 
the highest angular resolution, preferably obtained with the  VLBA
or the VLBI.
Based on multifrequency VLBA observations 
of some low-luminosity AGNs, Anderson et al. (2004) have shown
 that, even for the unresolved milli-arcsecond core,
 the spectra, luminosity, and size limits are consistent with 
 emission from jets. The
observed radio flux must therefore constitute only an upper limit
on the radio emission from the accretion flow itself. 

\subsection{UV Data}
By construction, all the objects in the sample have 
UV measurements obtained with HST by M05,
using the ACS in its High Resolution Camera (HRC) mode. M05
imaged each target in the
F250W band ($\lambda_{\rm central}\approx 2500$~\AA, FWHM$\approx 550$~\AA)
and in the
F330W band ($\lambda_{\rm central}\approx 3300$~\AA, FWHM$\approx 400$~\AA).
Between July 2002 and 
July 2003, 11 of the 13 LINERs were observed from
 two to five times each. Two objects,
NGC~404 and NGC~1052, were observed only once. 
As shown by M05, the image morphologies in all 13 objects are characterized 
by isolated, unresolved, nuclear UV sources, facilitating.
 the photometric measurements. As described in detail in M05,
 ACS/HRC photometry is very stable, with a photometric rms
scatter of less than 1 per cent in both the F250W and the F330W bands.

The M05 data permit extracting several measures of UV flux. The
variable UV flux provides a firm lower limit to the intrinsic, 
unextinguished, AGN UV flux. However, the highest observed flux,
to which I will refer as the ``high flux'', is also 
interesting; its non-variable fraction could be mainly or wholly 
from the AGN as well, but simply did not happen to vary during the M05
monitoring campaign. If there is some internal extinction by dust, the
actual flux could be even higher. On the other hand, some or
all of the non-variable fraction could be due to massive
stars, and could undergo little or no extinction, 
in which case the high flux 
would be an overestimate of the AGN flux.

In addition to the M05 data, I consider both the absolute and variable
UV fluxes of each object based on previous UV observations with HST,
as discussed below individually for each object. Most of these 
previous data are at $\sim 2300$~\AA, which for the present purposes
I will consider close enough in wavelength to be equivalent to the 2500~\AA~
measurements of M05. 

\subsection{X-ray Data}
X-ray data were compiled for most of the sample based on observations with 
{\it Chandra} and/or {\it Newton-XMM}, which have angular resolutions
of $\sim 1''$. In a few cases, even though such 
data exist, the nuclear sources are too bright to be measured reliably
by {\it Chandra}, due to ``pile-up'' in its detectors.
In these cases, measurements from older satellites are used, after
considering the evidence for the influence of non-nuclear sources on the
measured flux. When X-ray power-law photon spectral indices are not
available, I will assume a typical low-luminosity photon index of
$\Gamma=1.8$ (e.g., Ho et al. 2001), where the photon flux per energy 
interval, $d\epsilon$, is $dN/d\epsilon\propto \epsilon^{-\Gamma}$.   
 
\subsection{Distances} 
I will describe the SEDs of the AGNs in the sample 
in terms of their luminosities in the different bands, 
as calculated using an assumed distance to each galaxy.  
My analysis of interband luminosity ratios will not 
depend on the assumed distances, but
accurate distances can be important in some SED studies
(e.g., errors in distances can produce an artificial correlation
between luminosities in different bands).
I adopt the recent distance measurements to the sample galaxies 
compiled from the literature by M05. In 11/13 cases (the exceptions
are NGC~3486 and NGC~3642), the distances are
based on ``modern'' methods -- Cepheids, surface-brightness
fluctuations, tip of the red giant branch, and Tully-Fisher.

\subsection{Black Hole Masses}
Four of the objects in the sample (M81, NGC~3998, M87, and NGC~4594)
have in the literature 
measured BH masses based on stellar or gas kinematics, which I adopt.
In most of the remaining cases, I estimate BH masses using the 
Tremaine et al. (2002) mass vs. velocity-dispersion relation,
using stellar velocity dispersion measurements taken from the Hyperleda 
database\footnote{\tt http://leda.univ-lyon1.fr}.
An exception is NGC~4203, for which the Tremaine et al. (2002)
relation predicts a BH mass of 
$6.6\times 10^7 M_{\odot}$, but for which Sarzi et al. (2002)
set a $1\sigma$ upper limit of $2.4\times 10^7 M_{\odot}$. 
In this case I assume a mass of $1\times 10^7 M_{\odot}$. 

\subsection{Individual Objects}

\noindent{\bf NGC~404} --  Nagar et al. (2000) did not detect 
a radio core with the VLA in this galaxy at wavelengths of 0.7, 2, and 3.6~cm,
to $3\sigma$ limits of 3~mJy, 0.39~mJy, and 0.27~mJy, respectively. 

HST UV spectroscopy of the compact nucleus by 
Maoz et al. (1998) showed clear absorption signatures of OB stars,
contributing at least 40 per cent of the UV light. 
However, the relative shallowness of the absorptions meant that the 
light from massive stars was diluted by another component, comparable in flux, 
which could be
a featureless AGN continuum, or the light from less massive stars in 
an aging or continuous starburst.
In the single epoch of HST/ACS data by M05, the 
2500~\AA~ flux ($74\times 10^{-17}~\ergcmsA$) was $\approx 60$ per cent 
of the level measured by the 1994 spectroscopy 
analysed by Maoz et al. (1998; $115\times 10^{-17}~\ergcmsA$), 
and only 45 per cent of the HST Faint Object Camera (FOC) imaging 
measurement in 1993 
by Maoz et al. (1995; $180\times 10^{-17}~\ergcmsA$ at 2300~\AA). 
I will therefore adopt a lower limit on the nonstellar 2500\AA~
flux, of $(115-74)\times 10^{-17}~\ergcmsA$. At 2300\AA, the
AGN flux may have been as high as the level measured in 1993, 
minus the minimum stellar contribution, or approximately
$(180-0.4\times 115)\times 10^{-17}~\ergcmsA$, which I will adopt
as the ``high flux'' for this object. At 3300~\AA, M05 
measured a flux of $85\times 10^{-17}~\ergcmsA$,

In X-rays, a {\it Chandra} measurement of the compact nuclear source 
by Eracleous et al. (2002) 
gives an 0.5--2 keV flux of  $1.5\times10^{-14}$~\ergcms.

\noindent {\bf NGC~1052} --
Kadler et al. (2004) have used VLBA to resolve the core into
milliarcsecond-scale twin jets, with a total flux at   
0.7, 3.6, and 6~cm, of 
670~mJy, 2390~mJy, and 2410~mJy, respectively.
There is a small gap between the jets, and it is not clear
which component, if any, can be associated directly with the nucleus.

In the single epoch measured by M05 in 2002, the  
UV flux was $7.7\times 10^{-17}~\ergcmsA$ at 2500~\AA~ and 
$9.0\times 10^{-17}~\ergcmsA$ at 3300~\AA. The short wavelength
flux was at half the level of the 
2300~\AA~ flux in a 1997 HST Faint Object Spectrograph (FOS) 
spectrum (Gabel et al. 2000), as
measured by Pogge et al. (2000; $15\times 10^{-17}~\ergcmsA$),
which I will adopt as a high flux.
Based on the variable fraction, the lower limit on 
the AGN flux is then $(15-7.7)\times 10^{-17}~\ergcmsA$).

Kadler et al. (2004) have used {\it Chandra} to measure an unabsorbed nuclear  
0.2--8 keV flux of  $300\times10^{-14}$~\ergcms. They find the X-rays
are moderately absorbed by a column density of $\sim
10^{22}$~cm$^{-1}$. The photon index they derive, $\Gamma=0.3$, is
extremely low, but they note that this could be the result of pile-up 
 in the detector at low energies. I therefore adopt the photon index
of $\Gamma=1.4$ obtained by Guainazzi et al. (2000) using {\it BepposSAX}.
For this index, the 2--10 keV flux would be  $240\times10^{-14}$~\ergcms 

\noindent {\bf M81 (NGC 3031)} --
Bietenholz et al. (2000) used VLBI to measure in this galaxy 
a radio core
with 132~mJy at 3.6~cm, and Nagar et al. (2002) obtained
165~mJy at 2~cm with the VLBA.

M81 was imaged by M05 at five epochs. The mean level at 
2500\AA~, $200\times 10^{-17}~\ergcmsA$, was similar to
the one measured  by Maoz et al. (1998) at 1500~\AA~ 
in the 1993 HST/FOS spectrum of Ho et al. (1996)
  -- $150\times 10^{-17}~\ergcmsA$. [From an analysis of the FOS target
acquisition records, Maoz et al. (1998) deduced that M81 was 
located at the edges of its peak-up scans, possibly leading to some
light loss.]  The flux level meausured by M05
is also the same as
the 2200~\AA~ flux estimated by Maoz et al. (1998) by extrapolating
the 1996 WFPC2 measurement at $\sim 1600$~\AA~ by Devereux et al. (1997). 
The mean 3300~\AA flux found by M05 
was $130\times0.11\times 10^{-17}~\ergcmsA$.
Variations measured by 
M05 in both F250W (8 per cent) and F330W (11 per cent) 
give a lower limits on the AGN flux of
$200\times0.08\times 10^{-17}~\ergcmsA$, 
and 
$130\times0.11\times 10^{-17}~\ergcmsA$, respectively.

Ho et al. (2001) measured with {\it Chandra} an unabsorbed
2--10~keV flux
from the nucleus of $1.0\times 10^{-11}$\ergcms. Because
the nuclear source was heavily piled up, the counts were
estimated from the readout trail. 
  
A central BH mass of $6\times 10^7 M_{\odot}$ has been reported
by Bower et al. (2000) based on stellar kinematics, and
$7\times 10^7 M_{\odot}$ 
by Devereux et al. (2003) based on gas kinematics. I adopt the 
mean of the two.
 
\noindent {\bf NGC~3368} -- 
Nagar et al. (2000) report 
a $3\sigma$ limit of 0.6~mJy at 2~cm on any radio core
at VLA resolution ($\sim 1''$). 

MO5 found no UV variations
between the two epochs, in 2002 and 2003, at which this LINER~2 was observed.
However, the 2500~\AA~ flux ($22\times 10^{-17}~\ergcmsA$). 
was a factor of 4.5 higher than the 2300~\AA~
flux measured in 1993 with HST/FOC by Maoz et al. (1996;
$5\times 10^{-17}~\ergcmsA$). 
The variable fraction gives a lower limit on the AGN flux
of $(22-5)\times 10^{-17}~\ergcmsA$). 
For the high flux, I adopt $22\times 10^{-17}~\ergcmsA$ at 2500~\AA
and $30\times 10^{-17}~\ergcmsA$ at 3300~\AA.

In X-rays, Satyapal et al. (2004) used {\it Chandra} to measure
a 2--10~keV flux of $16\times10^{-14}$~\ergcms.
 
\noindent {\bf NGC~3486} -- 
In this bordeline Seyfert/LINER nucleus, no radio 
cores at 6~cm and 20~cm were detected at a VLA
resolution of $1''$ by Ho \& Ulvestad (2001), to a 
$3\sigma$ limit of 0.12~mJy~beam$^{-1}$.

In the UV monitoring by M05, this source displayed
neither short-term changes between the two, closely spaced (by 1
month), epochs in which it was observed, nor
long-term variations when comparing the 2500~\AA~ flux to a 
2300~\AA~ measurement in 1993 with the HST/FOC by Maoz et al. (1996).
The 2500~\AA~ and 3300~\AA~ flux levels are
$10.8\times 10^{-17}~\ergcmsA$, 
and $18.2\times 10^{-17}~\ergcmsA$, respectively.
Thus, there is no variability-based lower limit
on the AGN flux in this object. 

Ho et al. (2001) used {\it Chandra} to set an upper limit of 
$0.5\times10^{-14}$~\ergcms on the 2--10~keV nuclear flux.

Given the X-ray and radio non-detections and the UV non-variability,
M05 discussed the possibility that
this is another non-AGN LINER with UV emission dominated
by stars, like NGC~4569 and NGC~5055, which I have excluded from
the present sample for this reason. M05 noted, however,
that  M81 and M87, which are clearly AGNs with variable UV flux, 
were also near their ``historical'' UV levels in the M05 campaign,
and were constant in the two closely spaced epochs (for M87)
or in four out of five epochs (for M81). Thus, detection of
short-term variability in NGC~3486 might
 have been possible with better temporal sampling. 
I will therefore keep NGC~3486 in the sample, allowing for the 
possibility that its UV flux is AGN dominated. 
 
\noindent {\bf NGC~3642} -- 
Nagar et al. (2000) placed a $3\sigma$ upper limit of 0.39~mJy
on the 2~cm radio flux from this nucleus, at VLA resolution.
  
M05 measured 8 per cent peak-to-peak amplitude variations
around the mean 2500~\AA~ level, 
$24.5\times 10^{-17}~\ergcmsA$), providing a lower limit
on the AGN flux. The 3300~\AA~ flux was
$22.1\times 10^{-17}~\ergcmsA$), but not definitively
variable.  
The 30 per cent increase
in 2500~\AA~ flux compared to the 2300~\AA~ WFPC2 measurement in 1994 
by Barth et al. 
(1998; $19\times 10^{-17}~\ergcmsA$) 
was not deemed large enough to be considered significant, 
given the different bandpasses and
the UV sensitivity fluctuations of WFPC2. 
 
A {\it Rosat}-HRI measurement by Koratkar et al. (1995) shows
that the nuclear X-ray emission is concentrated within $\la 5''$.
Using the {\it ROSAT}-PSPC, Komossa et al. (1999) measured for this source 
a 1-2.4~keV flux of of $18\times10^{-14}$~\ergcms. 
For an assumed photon index of $\Gamma=1.8$, this would
correspond to a 0.5--2~keV flux of of $26\times10^{-14}$~\ergcms. 

\noindent {\bf NGC~3998} -- In this nucleus, Filho et
 al. (2002) have measured a variable radio core that is
unresolved at 5~mas resolution with VLBI, with a mean
6~cm flux of 83~mJy. 

In the UV, M05 measured in 2002--2003 a monotonic 20 per cent
decline in UV flux in  the F250W and F330W bands
(means of $199\times 10^{-17}~\ergcmsA$ and
$153\times 10^{-17}~\ergcmsA$, respectively), over the 11 months
they observed it. These variations provide firm lower
limits on the AGN flux. On long time-scales, 
the mean 2500~\AA\ flux level in 2003 
was about 5 times lower than reported by Fabbiano et al. (1994; 
$10^{-14}~\ergcmsA$) at 1740~\AA~ in 1992, based on FOC measurements. 
There is thus evidence for a large variable UV flux,
of order $10^{-14}~\ergcmsA$. However,
because of the different UV band, I will not adopt the Fabbiano
et al. (1994) point as a lower limit, and conservatively use
only the variable flux measured by M05.
Ptak et al. (2004) used the Optical Monitor on
{\it XMM-Newton} to roughly estimate a 2100~\AA~ UV flux of 
250 to $500\times 10^{-17}~\ergcmsA$ in 2001, intermediate to
the 1992 and 2003 levels.

The X-ray flux found by  Ptak et al. (2004) using 
{\it XMM-Newton} is 
$1100\times10^{-14}$~\ergcms at 2--10~keV, and is consistent with,
though perhaps a factor of $\sim 2$ higher than, previous X-ray 
measurements over the past two decades. The photon index is $\Gamma=1.9$.

A central BH mass of $2.7\times 10^8 M_{\odot}$ has been measured
by de Francesco et al. (2006) using gas kinematics.
 
\noindent {\bf NGC~4203} -- 
VLBA measurements by 
Anderson et al. (2004) of the unresolved (at the mas scale)
core provide the following fluxes: 9.9~mJy at 0.7~cm; 
                                   9.0~mJy at 1.35~cm; 
                                   10.2~mJy at 1.9~cm;
                                   8.5~mJy at 3.6~cm;
                                   8.1~mJy at 6~cm.   
Nagar et al. (2002) have found that the radio core is variable.

In the UV monitoring by M05, the nuclear source showed large fluctuations,
1.5 between maximum and minimum in F250W (mean: $58\times
10^{-17}~\ergcmsA$), and 1.4 in F330W (mean: $37\times 10^{-17}~\ergcmsA$).
 The 2500~\AA~ flux
level in 2003 was 3-4 times higher than in the HST/WFPC2 2300~\AA~ measurement
by Barth et al. (1998; $21\times 10^{-17}~\ergcmsA$), 
obtained in 1994. The variable UV flux, which provides a
lower limit on the AGN component, is thus $13\times
10^{-17}~\ergcmsA$ at 3300~\AA, and $(78-21)\times
10^{-17}~\ergcmsA$) at 2500~\AA. 

Ho et al. (2001) measured with {\it Chandra} a 2--10~keV flux
of $44\times10^{-14}$~\ergcms.
Terashima et al. (2002) measured with {\it ASCA} a  flux
of $119\times10^{-14}$~\ergcms at 0.5--2~keV, 
$205\times10^{-14}$~\ergcms 2--10~keV, and a photon index $\Gamma=1.8$. 
Terashima \& Wilson (2003)
verified that, although source was too bright 
to be measured with {\it Chandra}, the nuclear 
source is unresolved and dominates
the emission at the {\it ASCA} spatial resolution. 
I will adopt at 0.5--2~keV the {\it ASCA} measurement, and at 
2--10~keV the mean of the  {\it ASCA} and {\it Chandra} measurements, 
$125\times10^{-14}$~\ergcms, with a photon index $\Gamma=1.8$. 

As noted above, the Tremaine et al. (2002)
relation predicts, given the velocity dispersion 
in this galaxy, 167 km~s$^{-1}$, a BH mass of 
$6.6\times 10^7 M_{\odot}$, but Sarzi et al. (2002)
set an upper limit of $2.4\times 10^7 M_{\odot}$. 
In this case I will therefore assume a mass of $1\times 10^7 M_{\odot}$. 
  
\noindent {\bf M87 (NGC 4486)} -- 
A 2~cm VLBA measurement of the unresolved core of M87 by 
Kellerman  et al. (2004), with 1~mas resolution, 
is 3~Jy, consistent with a VLA ($1''$ resolution) measurement
by Biretta et al. (1991).
 
The UV flux is variable on short time-scales (Perlman et al. 2003; M05)
as well as on long ones (M05). At 2500~\AA, I will adopt
a high flux of  $100\times 10^{-17}~\ergcmsA$, based on an HST/FOC 
measurement by Maoz et al. (1996). A lower limit of the nonstellar
AGN flux in the UV is obtained from the difference between this
measurement and that of M05, $(100-55)\times 10^{-17}~\ergcmsA$.
At 3300~\AA, M05 found $48\times 10^{-17}~\ergcmsA$, and little 
variability between their two, closely spaced, epochs.

In X-rays,  Wilson and Yang (2002) measured 
with {\it Chandra} a flux density at 1~keV 
of $37\times10^{-14}$~\ergcms~keV$^{-1}$. 
Di Matteo et al. (2003) found, using the same data,
$(80\pm 2)\times10^{-14}$~\ergcms~keV$^{-1}$. 
I will adopt the mean of these two observations,
$59\times10^{-14}$~\ergcms~keV$^{-1}$. Both 
analyses find a photon index of $\Gamma=2.2$
 
A central BH mass of $3.4\times 10^9 M_{\odot}$ has been measured
by Macchetto et al. (1997) using gas kinematics.

\noindent {\bf NGC~4552} -- 
Nagar et al. (2002) measured with the VLBA at 6~cm (2~mas resolution)
a flux 99.5~mJy, and  with the VLA at 2~cm (150~mas resolution)
a flux of 59~mJy

The various UV measurements of this nucleus with HST are:
$1.5\times 10^{-17}~\ergcmsA$ (at 2300~\AA) and $1.8\times 10^{-17}~\ergcmsA$
(at 2800~\AA), with HST/FOC in 1993 (Cappellari et al. 1999);
$2\times 10^{-17}~\ergcmsA$ at 2500~\AA~
with HST/FOS in 1996 (Cappellari et al. 1999);
M05 observed it at two epochs between which it brightened by 20 per cent 
in both F250W and F330W. 
I adopt the means of the measurements of M05,
$2\times 10^{-17}~\ergcmsA$ (F250W), and
$1.5\times 10^{-17}~\ergcmsA$ (F330W),
and take 20 per cent of them as lower limits on the AGN flux.
  
Flohic et al. (2006) used {\it Chandra} to derive unabsorbed 
fluxes for the nuclear source  of $5.0\times 10^{-14}$\ergcms 
at 0.5--2~keV, and  $6.0\times 10^{-14}$\ergcms 
at 2--10~keV, and a photon index of $2.0\pm 0.2$.
Machacek et al. (2006) find a photon index of 1.7, and I will adopt
a value of 1.8.

\noindent {\bf NGC~4579} -- 
VLBA measurements by 
Anderson et al. (2004) of the radio core, unresolved
at 1~mas, are:                     14.6~mJy at 0.7~cm; 
                                   11.1~mJy at 1.35~cm; 
                                   15.3~mJy at 1.9~cm;
                                   16.2~mJy at 3.6~cm;
                                   19.2~mJy at 6~cm.

The various UV fluxes that haven been measured with HST, from low
to high, are:
at $\sim 2300$~\AA in 1994, 
$33\times 10^{-17}~\ergcmsA$ (Barth et al. 1996; Maoz et al. 1998);
at 2500~\AA~ in 2003,
$61\times 10^{-17}~\ergcmsA$ (M05, mean of two epochs);
at 3300~\AA~ in 2003,
$42\times 10^{-17}~\ergcmsA$ (M05, mean of two epochs);
and at $\sim 2300$~\AA~ in 1993, 
$110\times 10^{-17}~\ergcmsA$ (Maoz et al. 1995);     
 Between the two epochs of M05,
separated by less than a month, the nucleus brightened by 7 per cent 
in both UV bands.
I will adopt the Maoz et al. (1995) measurement as a high point
at 2500~\AA, 
the difference between the Maoz et al. (1995) and the Maoz et
al. (1998) measurements at 2200~\AA~ as lower limits on the AGN flux,
the M05 mean flux at 3300~\AA~ as a high point, and 7 per cent of this
value as the lower limit on the AGN flux at 3300~\AA.

Eracleous et al. (2002) find with {\it Chandra} an unabsorbed
 2--10~keV nuclear flux
of  $1300\times10^{-14}$~\ergcms, with a photon index of 1.8.
Analysis by Cappi et al. (2006)
of a measurement with {\it XMM-Newton}  gives 
$300\times10^{-14}$~\ergcms at 0.5--2~keV, and
$380\times10^{-14}$~\ergcms at 2--10~keV, with a photon index of 1.7.
Since pile-up is a concern in {\it Chandra} data for this bright
source (Terashima \& Wilson 2003), I will adopt the XMM values.

\noindent {\bf NGC~4594} -- 
The radio core flux at 6~cm measured by Hummel et al. (1984)
with the VLA at $1''$ resolution was 123~mJy, consistent with
a VLBI measurement by Graham et al. (1981).
At 2~cm, Hummel et al. (1984) measured an
unresolved ($< 0\farcs02$) flux of 100~mJy.

M05 reported short-term
UV variations, with 20 per cent peak-to-peak amplitude in F250W 
and 11 per cent in F330W, and 
mean UV fluxes of
$7.5\times 10^{-17}~\ergcmsA$ (F250W), and
$15.3\times 10^{-17}~\ergcmsA$ (F330W).
I take as a lower limit on the 2500~\AA~ AGN flux
the difference between the level
measured with the FOS in 1995 (Nicholson et al. 1998; 
Maoz et al. 1998) and the lowest level found by M05, 
$(12-6.9)\times 10^{-17}~\ergcmsA$, and a high flux
based on the FOS measurement.
A lower limit at 3300~AA, based on the variability observed by M05, is 
 $0.11\times 15.3\times 10^{-17}~\ergcmsA$.

In X-rays, I use the XMM-{\it Newton} measurements by 
Pellegrini et al. (2003b),
$55\times10^{-14}$~\ergcms~ at 0.5--2~keV, and
$130\times10^{-14}$~\ergcms~ at 2--10~keV, with $\Gamma=1.9$.

A central BH mass of $1\times 10^9 M_{\odot}$ has been measured
by Kormendy et al. (1996) using stellar kinematics.

\noindent {\bf NGC~4736} --  
VLA measurements with a resolution of $0\farcs 15$
reveal an unresolved nuclear source with a flux of 1.7~mJy at 2~cm
(Nagar et al. 2005). At 3.5~cm, with a resolution of
$0\farcs 24$,
K\"ording et al. (2005)
also find a flux of 1.7~mJy. 

Maoz et al. (1996) used the HST/FOC in 1993 to measure
a 2300~AA~ flux of
 at $19\times 10^{-17}~\ergcmsA$. M05 reported
that, in the ACS F250W band in 2003, 
the nucleus was significantly brighter than in 1993,
  at $48\times 10^{-17}~\ergcmsA$. 
However, reanalysis of the M05 images raises some doubts.
The nuclear source in this galaxy is superimposed on a fairly
bright diffuse stellar background, and the nuclear flux is sensitive
to the aperture used for the measurement. Compared to the FOC F220W
image, the stellar background
is more prominent in the redder F250W measurement by M05 which, being
a CCD observation, is also more susceptible to red leak. 
Photometry of the M05 data using an aperture smaller than 
used by M05 (and smaller than the minimum that M05 found was
required to provide reliable photometry in these ACS data) gives a value
closer to the 1993 level. On the other hand, the 1993 FOC data 
were affected by non-linearity and saturation, both of which would
lead to an underestimate of the true flux. Given these
uncertainties, I will adopt the mean level between these
measurements, $39\times 10^{-17}~\ergcmsA$ for the 2500~\AA~
flux, and the M05 value at 3300~\AA, $73\times 10^{-17}~\ergcmsA$.
The UV variability amplitude during the M05 campaign was only 5 per cent
at 2500~\AA, and null at 3300~\AA,,
providing a weak lower limit on the AGN flux. 

In {\it Chandra} X-ray images obtained
by Eracleous et al. (2002), the unresolved nuclear source (X-2 in their
table 4) has a 2--10~keV flux of 
$27\times10^{-14}$~\ergcms at 2--10~keV, and a photon index of 1.6.

\section{Derived Quantities}
\label{derived}

\begin{table*}
\begin{minipage}{\textwidth}
\begin{tabular}{l|r|c|c|r|c|r|c|c|c}
\hline
\hline
{Object} &
{$\log(\nu L_\nu)$} &
{$\log(\nu L_\nu)$} &
{$\log(\nu L_\nu)$} &
{$\log(\nu L_\nu)$} &
{$\log R_{\rm UV}$} &
{$\log R_{\rm UV}$} &
{$\alpha_{\rm ox}$} &
{$\alpha_{\rm ox}$} &
{$\log(\nu L_{\nu}/L_{\rm E})$} \\
{} &
{6~cm} &
{2500~\AA~l.l.} &
{2500~\AA} &
{2~keV} &
{u.l.} &
{} &
{u.l} &
{} &
{2500~\AA} \\
{(1)} &
{(2)} &
{(3)} &
{(4)} &
{(5)} &
{(6)} &
{(7)} &
{(8)} &
{(9)} &
{(10)}\\
\hline 
NGC~~404& $<$ 34.44&$>$  39.24&  39.75&  37.14&$<$ 0.59&$<$ 0.08&$<-$1.81&$-$2.00 & $-$3.7     \\  
NGC~1052&  39.68&$>$  39.93&  40.24&  40.54&$<$  5.13&  4.81&$<-$0.77&$-$0.89	      & $-$5.9\\
   M81&  37.21&$>$  39.04&  40.14&  39.92&$<$  3.55&  2.45&$<-$0.66&$-$1.08	      & $-$5.8\\
NGC~3368& $<$ 35.71&$>$  39.84&  39.95&  39.06&$<$ 1.25& $<$ 1.14&$<-$1.30&$-$1.34  & $-$5.5\\
NGC~3486& $<$ 34.60& \nodata&  39.31& $<$ 37.24& \nodata&  $<$ 0.66&\nodata&$-$1.80 & $-$5.0\\
NGC~3642& $<$ 36.35&$>$  39.69&  40.78&  40.29&$<$ 2.05&  $<$ 0.96&$<-$0.77&$-$1.19 & $-$4.4\\
NGC~3998&  37.93&$>$  40.34&  41.06&  41.12&$<$  2.98&  2.26&$<-$0.70&$-$0.98	      & $-$5.5\\
NGC~4203&  37.05&$>$  40.63&  40.64&  40.35&$<$  1.80&  1.79&$<-$1.11&$-$1.11	      & $-$4.5\\
   M87&  39.73&$>$  40.57&  40.92&  40.02&$<$  4.54&  4.19&$<-$1.21&$-$1.35	      & $-$6.7\\
NGC~4552&  38.15&$>$  38.58&  39.27&  39.01&$<$  4.96&  4.26&$<-$0.83&$-$1.10	      & $-$7.4\\
NGC~4579&  37.71&$>$  41.13&  41.29&  41.07&$<$  1.96&  1.80&$<-$1.02&$-$1.08	      & $-$4.6\\
NGC~4594&  37.78&$>$  39.18&  39.56&  39.76&$<$  3.98&  3.61&$<-$0.78&$-$0.92	      & $-$7.6\\
NGC~4736&  35.48&$>$  38.21&  39.50&  38.54&$<$  2.66&  1.37&$<-$0.87&$-$1.37	      & $-$5.7\\
\hline
\end{tabular}
\end{minipage}
\caption{Derived Quantities. See \S\ref{derived} for details of quantity
  derivations. 
Column header explanations: (2)-(5) - log of the luminosity, $(\nu
L_{\nu})/({\rm erg~s}^{-1})$; (3) - lower limit of the 2500~\AA~
luminosity, based on the variable flux in Table~\ref{table1};
(6)-(7) - radio loudness parameter, $R_{\rm UV}\equiv L_\nu(6~{\rm
  cm})/L_\nu(2500~{\rm \AA})$; (6) - upper limit on $R_{\rm UV}$
based on the lower limit on the 2500~\AA~ luminosity;
(8)-(9) - $\alpha_{\rm ox}$ parameter, $\equiv 0.384 \log[L_\nu(2~{\rm
  keV})/L_\nu(2500~{\rm \AA})]$; (8) - upper limit on $\alpha_{\rm ox}$
based on the lower limit on the 2500~\AA~ luminosity;
(10) - log of the 2500~\AA~luminosity, $\nu
L_{\nu}$, as a fraction of the Eddington luminosity,  
$L_{\rm E}=1.3\times 10^{38} (M/M_{\odot})$.}
\label{table2}
\end{table*}

The analysis that follows will be based mainly on luminosities
at 6~cm, 2500~\AA, and 2~keV. The following conversions are
therefore performed on the data adopted in the previous section,
and listed in Table~\ref{table1}.

When high-angular-resolution 5~GHz fluxes are unavailable,
I assume
the median spectral index $s$ ($f_\nu\propto \nu^s$, where $\nu$ 
is frequency, and $f_\nu$ is monochromatic flux) found by
Nagar et al. (2001) for the core emission in a sample of low-luminosity 
AGNs, $s=-0.2$, in order to convert monochromatic
fluxes from 15~GHz to 5~GHz.

I correct all UV fluxes
for Galactic extinction, 
assuming the $B$-band extinction values (listed in Table~\ref{table1})
of Schlegel et al. (1998),
and the Galactic extinction curve of Cardelli et al. (1989), with
the parameter $R_V=3.1$.

Monochromatic X-ray fluxes at 2~keV are recovered from the 
0.5--2~keV or 2--10~keV fluxes listed in Table~\ref{table1}, for the 
adopted power-law photon spectral indices. When photon indices are not
available, I assume a typical low-luminosity photon index of
$\Gamma=1.8$ (e.g., Ho et al. 2001). When both 
0.5--2~KeV and 2--10~keV fluxes are available, I use the mean
of the two 2~keV monochromatic fluxes obtained.

All monochromatic fluxes, $f_\nu$, are converted to 
luminosities, $L_\nu$ and $\nu L_\nu$, using the adopted 
distances to the galaxies, listed in Table~\ref{table1}. Luminosities 
in units of the Eddington luminosity, 
$L_E=1.3\times 10^{38} (M/M_{\odot})$, are obtained
using the BH masses in Table~\ref{table1}.

The ratio of UV to X-ray luminosity in AGNs is usually 
discussed in terms of $\alpha_{\rm ox}$, 
the spectral index of a hypothetical power-law
between $L_\nu$ at 2500~\AA~ and at 2~keV, or
$$
\alpha_{\rm ox}\equiv\frac{\log[L_\nu(2500~{\rm \AA})/L_\nu(2~{\rm keV})]}
                     {\log[(\nu(2500~{\rm \AA})/\nu(2~{\rm keV})]} 
$$
\begin{equation}
 =0.384~\log[L_\nu(2~{\rm keV})/L_\nu(2500~{\rm \AA})].
\end{equation}
 I derive two 
values of $\alpha_{\rm ox}$ for every galaxy in the sample,
one based on the ``high point'' UV flux, and one based
on the lower limit on the UV flux from an AGN, based on the
variable flux. The latter provides an upper limit on 
$\alpha_{\rm ox}$. One galaxy, NGC~3486, did not vary 
during the M05 campaign or before it, and hence there is no lower limit 
on its AGN UV flux. Furthermore, the nucleus is undetected    
in X-rays with only an upper limit. In this case I therefore use
the UV ``high point'' and the X-ray upper limit to derive
only an upper limit on $\alpha_{\rm ox}$.
  
``Radio loudness'', $R$, is usually discussed in terms of the ratio
of the luminosity at 5~GHz  to the luminosity in optical, UV, or X-ray 
bands. I will define $R_{\rm UV}$ as the ratio of $L_\nu$
between 5~GHz and 2500~\AA,
\begin{equation}
R_{\rm UV}\equiv L_\nu(6~{\rm
  cm})/L_\nu(2500~{\rm \AA}).
\end{equation}
 As in the case of $\alpha_{\rm ox}$,
I calculate two values of $R_{\rm UV}$ for each galaxy, one based on the high
UV measurement in Table~\ref{table1}, and another based on the lower limit
on the nonstellar UV flux, which gives an upper limit on $R_{\rm UV}$.
In three galaxies, NGC~404, NGC~3368, and NGC~3642, no radio core
has been detected. The upper limit on the radio flux in each of these
cases, combined with the UV high point and lower limit, give two
separate upper limits on $R_{\rm UV}$. In a fourth case, NGC~3486, there
is an upper limit on the radio flux, and no UV variability is detected.
In this case, there is only one upper limit on $R_{\rm UV}$, based on the 
constant UV flux.  

\section{Results}

The numbers compiled and derived above permit a renewed look at the 
SEDs of unobscured, LINER-type, low-luminosity AGNs, particularly the ratios
of their UV, X-ray, and radio luminosities.
Table~\ref{table2} lists the monochromatic luminosities $\nu L_\nu$ 
of every object at the various frequencies,
based on the fluxes in Table~\ref{table1} in the different bands, after
the necessary corrections and conversions, and
 the main derived quantities for the sample:
$\alpha_{\rm ox}$, and $R_{\rm UV}$, each based on
both the lower-limit UV flux and the high UV flux.

\subsection{The SED}

\begin{figure*}
\includegraphics[width=0.90\textwidth]{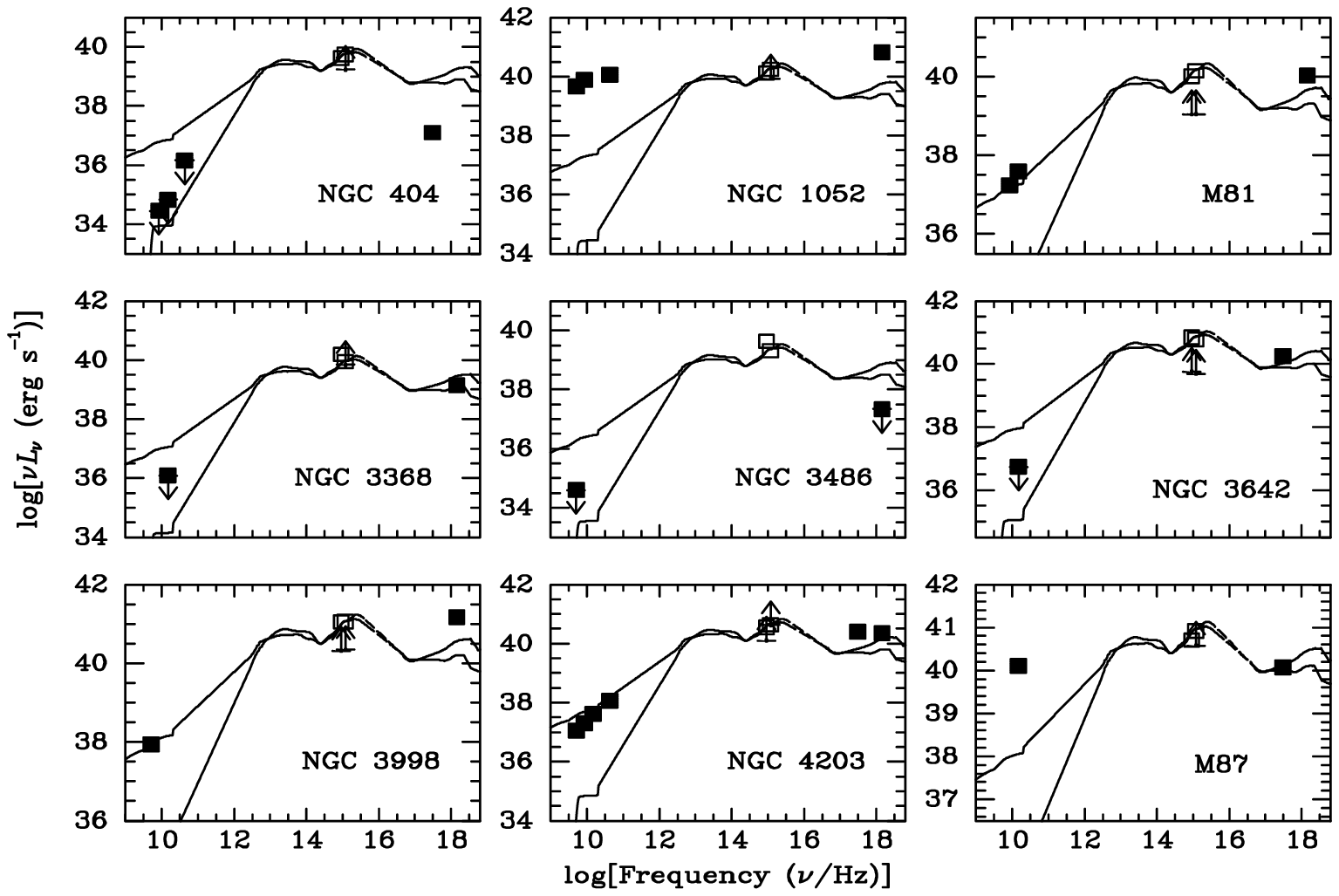}
\end{figure*}
\begin{figure*}
\includegraphics[width=0.90\textwidth]{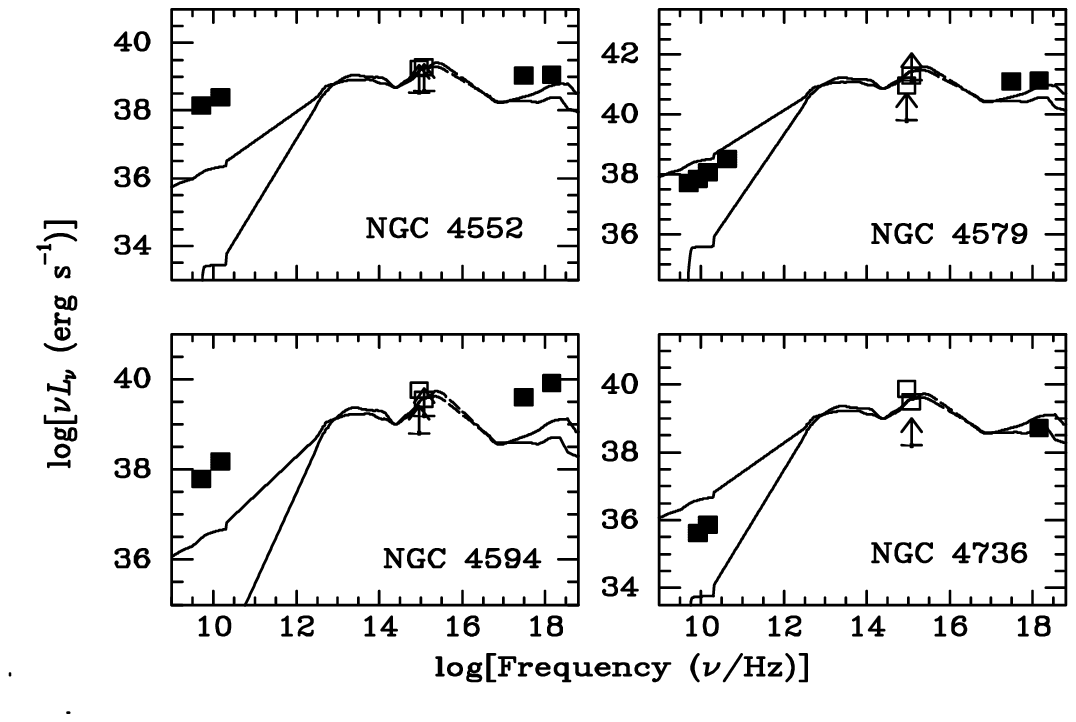}
\caption{Radio through X-ray spectral energy distributions for the
13 LINERs, in $\log \nu L_{\nu}$ vs. $\log \nu$.
Upper limits are $3\sigma$. Lower limits in the UV are based
on the variable, and hence nonstellar, flux. The solid curves show the 
mean SEDs of radio-loud and radio-quiet quasars from Elvis et
al. (1994), normalized to pass through the high UV 2500~\AA~
measurement of each LINER.
}
\label{sedfig}
\end{figure*}

Figure~\ref{sedfig} displays the SED data of each object, based
on the data in Table~\ref{table1}. 
Following Ho (1999), I overlay in every frame the
mean SEDs of radio-loud and radio-quiet quasars from Elvis et
al. (1994), normalized to pass through the high UV 2500~\AA~
measurements for the LINERs. All the multi-wavelength 
measurements I use are
non-simultaneous, often with many years between the measurements
in different bands. Variations by factors of a few in each band
are common over these time-scales in low-luminosity objects (see, e.g.,
M05, and references therein). This uncertainty should be kept 
in mind when comparing the non-simultaneous measurements of 
each individual object to the mean quasar SEDs.

The sample of LINERs discussed here has been selected to be 
unobscured, is the sense that a nuclear UV point source is detected,
optical imaging, when available,
 has not shown evidence for foreground dust extinction,
and X-ray absorbing columns are $N_H\la 10^{22}$~cm$^{-2}$.
 From Fig.~\ref{sedfig},
it is qualitatively evident that the SEDs of this sample are 
not dramatically different from those of quasars [the typical
quasars used to produce the Elvis et al. (1994) templates have $\nu
L_{\nu}(2500~{\rm \AA})\sim 10^{45}~{\rm erg~s}^{-1}$]. While
the ratio of X-ray to UV luminosity is sometimes larger in 
the LINERs, the difference is by not by more than a factor of a few.
Furthermore, there is no clear evidence for the absence of 
an optical-UV bump, only perhaps some signs that such a bump
may be weaker, relative to X-rays, compared to quasars. 

In the radio, Fig.~\ref{sedfig} confirms previous assessments that, compared
to quasars, most low-luminosity AGNs are radio loud, or 
even ``super radio loud''. The radio loudness and UV-to-X ratio
are examined more quantitatively below. 

\begin{figure*}
\includegraphics[width=0.85\textwidth]{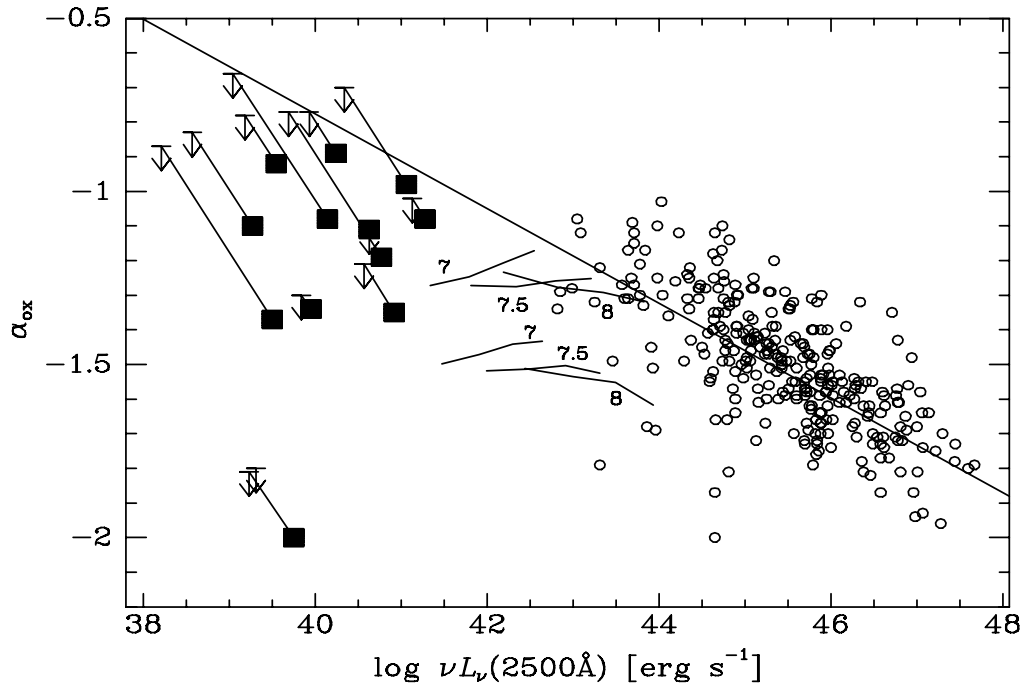}
\caption{The UV-to-X-ray spectral index, 
$\alpha_{\rm ox}$, vs. 2500~\AA~ luminosity, $\nu L_\nu (2500~{\rm \AA})$.
For every LINER, 
the  value obtained
by using its high UV point is shown with a filled square,
the value based on its UV lower limit is marked
as an upper limit on $\alpha_{\rm ox}$, and the two are connected with a line. 
Small symbols reproduce the compilation by Steffen et al. (2006) of
$\alpha_{\rm ox}$ and 2500~\AA~ luminosity 
for several samples of broad-lined AGNs. 
The best-fitting trend between these variables found by 
these authors for their sample is also shown (straight line).  
The $\alpha_{\rm ox}$ values
for most of the LINER sample span the range $-1.4<\alpha_{\rm
  ox}<-0.8$, largely overlapping with
 that of the Seyferts in the Steffen et al. sample.  
The short curves are from models by Merloni \& Fabian (2002) of coronal
outflow dominated thin accretion discs (see \S\ref{discuss}). Each
curve is labelled by its black hole mass, $\log (M_{\rm
  BH}/M_{\odot})$. The accretion rate along each curve
decreases from $10^{-2}L_E$ on the right end to $10^{-3.5}L_E$ on the
left. Models are shown for two values of the numerical factor $K$,  
which contains the unknown quantities in the model, and is 
related to the maximal fraction of power released in the corona.
The top three curves are for $K=0.95$ and the bottom three are for
$K=0.775$.
}
\label{alphaoxfig}
\end{figure*}

\subsection{$\alpha_{\rm ox}$}

Figure~\ref{alphaoxfig} 
permits a more quantitative look at the UV-to-X ratio
by showing $\alpha_{\rm ox}$ as a function of UV luminosity.
For every object in the LINER sample, 
I plot the two $\alpha_{\rm ox}$ values obtained
by using either the high UV point or the UV lower limit, and connect
them with a line. The true $\alpha_{\rm ox}$ is therefore somewhere
along these connecting lines, but possibly even below the point
corresponding to the highest UV measurement, because UV measurements
are so susceptible to extinction. We thus see that $\alpha_{\rm ox}$
for most of the sample spans the range $-1.4<\alpha_{\rm ox}<-0.8$.  
  
To place these results in the wider AGN context, I reproduce
in Fig.~\ref{alphaoxfig} the compilation by Steffen et al. (2006) of
$\alpha_{\rm ox}$ and 2500~\AA~ luminosity 
for several samples of broad-lined AGNs. Also plotted (straight line)
is the best-fitting trend between these variables. as found by 
Steffen et al. (2006) for their sample. While the $\alpha_{\rm ox}$
values for the LINERs are high compared to those of luminous
quasars, as we saw quantitatively in Fig.~\ref{sedfig} (most of the quasars
in the Elvis et al. 1994 sample have $-1.5<\alpha_{\rm ox}<-1.2$), 
they are actually 
{\it in a similar range to those of Seyfert 1 galaxies}, 
having luminosities up to $\sim 10^{44}$\ergs.
The $\alpha_{\rm ox}$'s of the LINERs are clearly below the 
extrapolation of the Steffen et al. relation to low luminosities.
Indeed, based on their data alone, Steffen et al. already noted
evidence (significant at the $2\sigma$ level) for a flattening
of the relation toward low luminosities. Figure~\ref{alphaoxfig} shows that
such a turnover is unavoidable. 
Similarly, the  $\alpha_{\rm ox}$'s of the LINERs in the present
sample are very similar to the values found by Greene \& Ho (2007), $-1.0$
to $-1.2$, for a sample of type-1 AGNs having intermediate-mass BHs
($\sim 10^{5-6}M_\odot$), with luminosities
$\nu L_\nu(2500~{\rm AA})=10^{41-43}$~erg~s$^{-1}$. 
We thus see that unobscured low-luminosity LINER
AGNs, as far as their UV-to-X luminosity ratios are concerned, 
are quite similar to AGNs that are $10^4$ times more luminous
in UV and X-rays. 

Two galaxies, NGC~404 and NGC~3486, are very faint or undetected
in X-rays, leading to outlying low values of $\alpha_{\rm ox}$.
Interestingly, both galaxies have small expected BH masses, of
$\la 10^6 M_{\odot}$, an order of magnitude or more below those
of the rest of the sample. A larger sample of galaxies 
is required to see if such a trend, of underluminous X-ray sources
from small bulges, is real. One must keep in mind also the possibility
that both of these objects are not true AGNs, and that the nuclear UV
sources are, instead, compact star clusters. Maoz et al. (1998) indeed
found that at least some of the UV light in NGC~404 is from massive
stars. Although M05 found that 
the UV flux in NGC~404 differed significantly between their single
epoch and previous HST observations, there is always some risk in
such comparisons among measurements made with different instrumental
setups. As noted above, NGC~3486 has not been seen to vary on either
long or short time-scales, and so a nonstellar nature of its UV
emission has not been demonstrated. (In fact, since it is also 
undetected in radio, its optical 
 spectrum, showing a borderline LINER/Seyfert 2 nucleus, 
is the only indicator of a possible AGN.) Further data
are thus needed in order to determine if such objects are 
powered by stars, or constitute another AGN phase, characterized
by very steep $\alpha_{\rm ox}$, and perhaps limited to 
galaxies with small BHs.      

\subsection{Radio Loudness}

\begin{figure*}
\includegraphics[width=0.75\textwidth]{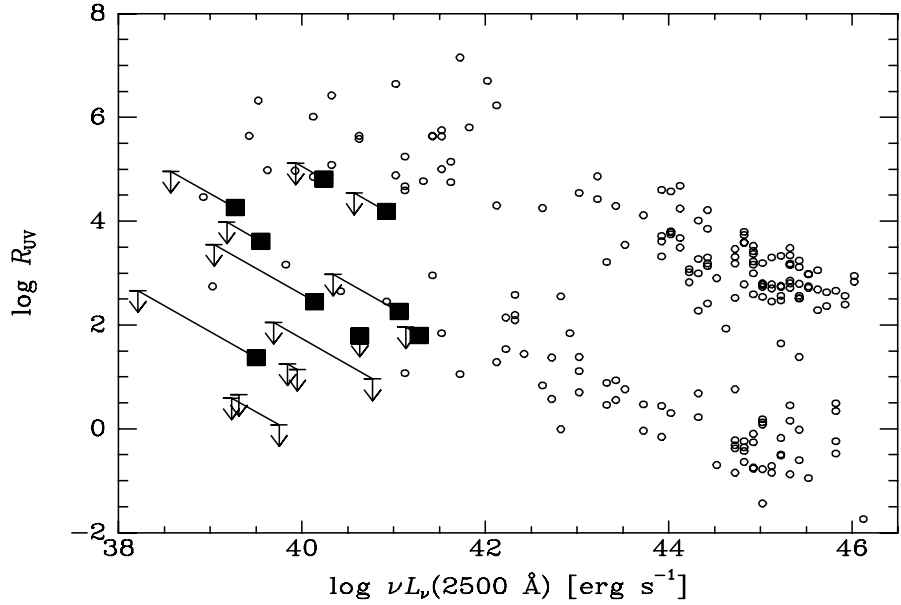}
\caption{The radio-to-UV (5~GHz to 2500~\AA) 
radio loudness parameter $R_{\rm UV}$, plotted
 vs. 2500~\AA~luminosity, $\nu L_\nu (2500~{\rm \AA})$.
As in Fig~\ref{alphaoxfig}, $R_{\rm UV}$ using 
each LINER's high UV point is shown with a filled square,
$R_{\rm UV}$ based on the UV lower limit is marked
as an upper limit, and the two are connected with a line. 
Double connected upper limits are based on the two UV measurements
for objects that are undetected in radio. The single upper limit
is NGC~3486, which is undetected in radio, and has no
variability-based UV lower limit.
Small symbols reproduce the compilation by Sikora et al. (2007) of
$R$ and $L_B$ for several samples of AGNs, 
after converting their values from 4400~\AA~ to
2500~\AA. 
While all low-luminosity AGNs are, on average, radio-louder by a 
factor $\sim 100$ than high-luminosity quasars, most of the LINERs
are actually on the ``radio-quiet'' branch of the distribution 
at low luminosities.
}
\label{RvLfig}
\end{figure*}

\begin{figure*}
\includegraphics[width=0.75\textwidth]{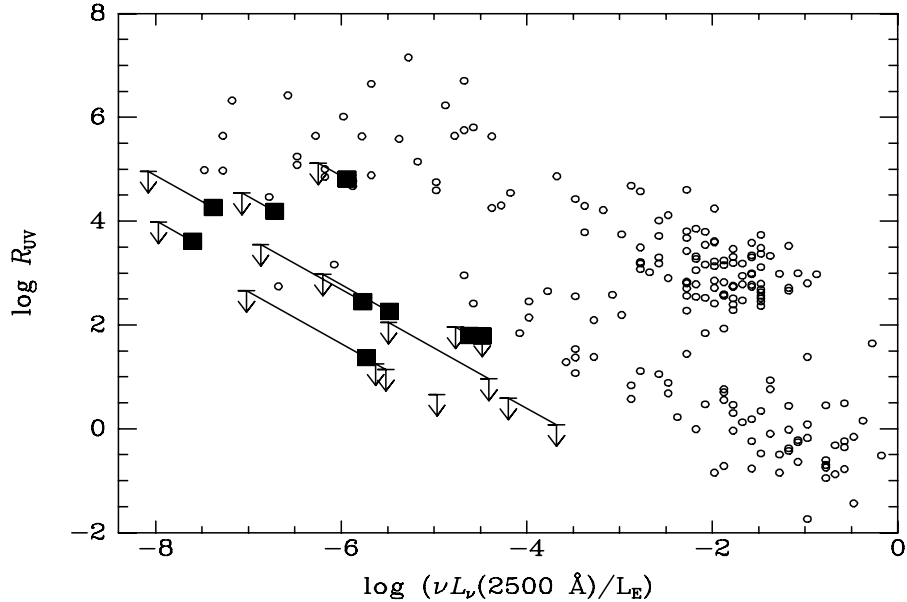}
\caption{Same as Fig.~\ref{RvLfig}, but normalizing the UV luminosity
of each object by it Eddington lumunosity, $L_E$.
}
\label{RvLeddfig}
\end{figure*}

Figure~\ref{RvLfig} plots the radio-loudness parameter, $R_{\rm UV}$, 
vs. UV luminosity
for the LINER sample. As in Fig.~\ref{alphaoxfig}, the two possible
values, based on the two UV measurements, are shown for
every galaxy, and are connected by lines. For comparison, I include
in Fig.~\ref{RvLfig} data from the recent compilation by Sikora et al. (2007) 
for several AGN samples. I convert the luminosities and $R$
parameters, which are given by Sikora et al. 
for the $B$-band (4400~\AA), to 2500~\AA,
assuming an optical-UV power-law relation 
$f_\nu\propto \nu^{-0.5}$.
The Sikora et al. (2007) points for the four LINERs 
in common with my sample (M81, NGC~3998, NGC~4203, NGC~4579) 
are not plotted. 
Figure~\ref{RvLeddfig} is the same as Figure~\ref{RvLfig}, except
that the two values of the UV luminosity for each object are
normalized by the object's Eddington luminosity.
(Note that Sikora et al., in their
fig.~4, plot $R$ only versus the Eddington ratio, $L/L_E$,
rather than $L$. However, as seen by comparing the Sikora et al. data
in my Figs.~\ref{RvLfig} and \ref{RvLeddfig},
the plot is hardly changed when using either $L$ or
$L/L_E$, the result of the relatively small range in BH
masses, compared to the large range in luminosities in their sample.)

Sikora et al. (2007) noted that, as pointed out by Ho (2002),
radio loudness in inversely correlated with Eddington ratio.
They showed, however, that the entire (possibly bimodal) 
distribution in $R_{\rm UV}$, including
both radio loud and radio quiet objects, shifts to higher $R$ values
when going from high to low luminosities. The radio-loud ``branch''
in the distribution
consists exclusively of massive elliptical galaxies, with correspondingly
large BHs, while the radio-quiet branch included both spirals and
ellipticals. Related results have been shown by Xu et al. (1999), 
Laor (2001), Best et al. (2005), Chiaberge et al. (2005),
 Wang, Wu, \& Kong (2006), Panessa
et al. (2007), and Chiaberge (2007).  

The LINERs in the present sample conform with this picture.
While their radio-to-UV luminosity ratios are often similar
to those of radio-loud quasars, and sometimes even greater
(as already seen qualitatively in Fig.~\ref{sedfig}), these AGNs 
fall on the same two branches on the diagram, with the
majority actually being on the ``radio-quiet'' branch, with $\log
R_{\rm uv}\approx 2$. 
(I note that, in their compilation of radio fluxes, Sikora et
al. 2007 included extended radio flux, while I have made a point 
of isolating, at the highest spatial resolution possible, just 
the unresolved nuclear flux. Inclusion of the extended flux in
the LINER sample would likely move some of the points upwards in
the diagram to some degree. This shift would be by up to an order
of magnitude for the objects on the ``radio-loud'' branch, e.g., M87,
which is a FRI radio galaxy, but probably not by much for the others, since
they are all core-dominated radio sources.) 
The four LINERs that are on the ``radio-loud branch'', with $\log
R_{\rm uv}\approx 4$, (NGC~1052,
M87, NGC~4552, and NGC~4594) indeed all have high BH masses, 
above $10^8 M_{\odot}$, and two of them have masses $>10^9 M_{\odot}$.
Equivalently, all the objects on the radio-loud branch have 
the lowest Eddington ratios, $\nu L_\nu (2500~{\rm \AA})/L_E\la 10^{-6}$.

Thus, in terms of their ratios of radio to UV luminosities, low-luminosity 
LINERs are, again, similar to AGNs of high luminosity, in that their
radio loudness spans about 4 orders of magnitude, most of them 
are at the low-$R$ end of the distribution, and the most radio-loud
cases occur in massive early-type galaxies.

\subsection{Comparison to Previous Work}
\label{comparisonsection}
The conclusions above, that LINER SEDs are overall similar to those
of higher-luminosity AGNs, are in contrast to those of most previous LINER
SED studies (see \S~1). It is therefore instructive to understand
the source of these different conclusions. 

The claims by previous authors 
for a distinct SED in low-luminosity AGNs, with a weak or absent
big blue bump, has been
based on: (1) radio loudness; (2) a low UV/X-ray ratio;
and (3) a steep optical-UV slope. In this paper, I have argued that (1)
and (2) are actually quite similar in AGNs at low and intermediate
luminosities. In terms of measurements, 
the values of $\alpha_{\rm ox}$, e.g., for the five objects
in common to Ho (1999), Ho et al. (2000), and 
to this work (M81, M87, NGC~4579, NGC~4594)
are similar. Thus,
the X-ray fluxes used by those authors, which were based 
on {\it Einstein}, {\it Rosat}, and {\it ASCA} 
measurements having lower angular resolutions,
did not significantly overestimate the AGN flux (due, e.g., 
to inclusion of diffuse X-ray emission or discrete circumnuclear
sources), and hence this is not the source of the discrepancy.
Rather, with the more recent data on the statistics of
$\alpha_{\rm ox}$ (e.g., Steffen et al. 2006; Greene \& Ho 2007)
and $R$ (e.g., Sikora et al. 2007), the values for LINERs are seen
to largely overlap with those for Seyferts. 

However, the main source of the discrepancy concerns
(3), the optical-UV slope of the SED, which I have chosen to 
ignore in the present work. 
Typical optical-UV power-law indices in quasars (assuming $f_\nu\propto
\nu^{\alpha_{\rm ou}}$) are $\alpha_{\rm ou}\approx -0.5$ 
(Shang et al. 2005), $-0.65$ (vanden Berk et al. 2001, at low
redshifts), or $-1$ (Zheng
et al. 1997). The exact value depends
on the chosen wavelength range (optical contamination is
a problem even in quasars), the bands to which the power law
is fit,
and how far to the UV one looks (the spectrum is not a pure power law,
and it becomes steeper toward the far UV). Furthermore the spectral slope
may depend on luminosity. By comparison, among LINERs that are apparently
unobscured, previous SED studies have measured
typical optical-UV slopes of  $\alpha_{\rm ou}\approx -1.5$
(although some LINERs, e.g., NGC~4579, 
have spectra that actually flatten
in the UV, to $\alpha_{\rm ou}\approx -0.5$; Maoz et al. 1998).
Is this difference in $\alpha_{\rm ou}$ between LINERs and
higher-luminosity AGNs significant?
 
Efforts to isolate the nonstellar optical continuum in LINERs
have been based on imaging (e.g., Chiaberge et al. 2006)
or on the
dilution of stellar features in optical spectra (e.g., Ho et
al. 2000). All of these attempts,
however, extrapolated the surface brightness outside the nucleus
inwards in order to subtract the starlight, and/or assumed an
unchanging stellar population when going from the bulge to the nucleus
at HST resolution. Many galaxies (including
the Milky Way), host compact nuclear star clusters. While those
clusters that are detected as such are often
young, in some galaxies the clusters could be of intermediate age, 
and hence much harder to discern as such, photometrically or
spectroscopically. Due to such residual contamination by starlight,
systematic errors by factors of a
few in the optical luminosity are conceivable. While such errors would
have little effect on estimates of 
radio loudness and $\alpha_{\rm ox}$, they would have a strong impact
on the optical-UV slope. 

Some of the previous studies have used spectral slopes measured 
in the space-UV region, where stellar contamination is less of a
problem. However, these estimates are extremely sensitive
to small amounts of foreground extinction. A $V$-band extinction of 
$A_V=0.2$~mag is sufficient to change a UV power-law
slope of $-0.5$ to $-1.5$ (see discussion 
in Maoz et al. 1998). The preponderance of lanes, wisps, and clumps of
dust seen in the neighbourhoods of most LINERs (see, e.g., Pogge et
al. 2000; Chiaberge et al. 2006) makes it likely that even 
relatively unobscured objects undergo some small degree of reddening.
Such reddening, rather than an intrinsic lack of UV emission,
 could be the cause of the UV steepening of LINER SEDs. 
(NGC~4579, with its UV slope of $-0.5$, may be the case of an
unobstructed line of sight.)
Given these uncertainties, I have ignored in this work measurements
of optical-UV slopes in LINERs. 
The remaining observables suggest a similarity, rather than a
difference, between the SEDs of LINERs and those of more luminous AGNs.

\section{Discussion and Conclusions}
\label{discuss}

The optical-to-UV emission in luminous AGNs is widely thought 
to come from the inner parts of a thermally radiating, geometrically
thin, accretion disc. If so, this emission provides the most direct
measure of the accretion rate on to the BH. Tracking this measure
to lower and lower luminosities requires reliance on the UV, which 
is less susceptible than the optical emission to contamination by the 
old stellar population of galaxy bulges.

I have used new, high-angular-resolution data, particularly recent
measurements and variability-based lower limits in the UV, to 
re-assess the SED of low-luminosity LINER nuclei. I have 
focused on unobscured objects in which the faint nuclear
emission has been properly isolated from surrounding contamination. 
I have ignored IR and optical data and optical-UV slopes, which can be
strongly affected by stellar contamination and by small amounts of
reddening by dust. With these choices, I have shown that 
the SEDs of LINERs are similar to those of Seyfert-type AGNs $10^2$ to
$10^4$ more luminous. 
This similarity is seen quantitatively in the parameters
$\alpha_{\rm ox}$ and $R_{\rm UV}$. The lack of any conspicuous ``phase
transition'' as a function of luminosity or accretion rate 
in the SEDs of LINERs, extends similar recent results by Panessa et
al. (2007) for low-luminosity Seyferts. 

It is tempting to speculate,
therefore, that the same combination of physical components that 
gives rise to the SEDs of quasars is present at luminosities and
accretion rates that are $\sim 10^{5-8}$ times smaller. In particular,
if radiatively efficient accretion discs produce the UV emission in
quasars, there is no compelling 
evidence that such discs disappear at low luminosities.   
Instead, the ratio of UV to X-ray emission is fairly insensitive to
luminosity. It is only the ratio of luminosities between  radio and other
wavelengths that does increase dramatically as the luminosity
decreases. However, it seems that the entire distribution of radio 
loudness shifts to higher values, with most objects 
remaining at the low side of the $R$ distribution. The decrease
in the accretion rate on to a supermassive BH (as traced by the UV
luminosity) could thus be manifest 
as a hand-in-hand decrease in the UV and X-ray luminosities, but with
a much smaller decrease in radio luminosity. 
The sources of the UV
radiation (presumably a thin accretion disc, although synchrotron
emission from the jet could also contribute or dominate, e.g., 
Chiaberge et al. 1999; Verdoes Kleijn et al. 2002) and of the X-rays
apparently persist, but are simply scaled down, with a minor increase
in the prominence of the X-ray emission.

Interestingly,
analogous results have been found recently for stellar-mass Galactic 
BHs. Miller et al. (2006a,b) have analysed data for three 
BHs (including a re-analysis of {\it ASCA} data for Cygnus X-1
in its low state) accreting at $\sim 10^{-2}$ to $\sim 10^{-3}$ 
of the Eddington rate. In each case,
the soft X-ray spectra require the presence of 
a thermally radiating thin accretion disc, down to the innermost marginally
stable circular orbit. There is no obvious reason why such ``mass-starved''
structures could not exist at even-lower low accretion rates.
RIAFs were devised in order to explain the low
luminosities that are observed from dormant galactic nuclei,
despite the significant rates of mass infall
expected on to the BHs. Instead, the evidence for the persistence of thin,
radiatively efficient, but mass-flow-starved, discs suggests that
some mechanism prevents gas from reaching the inner parts of the 
accretion flow in the first place. The radio loudness at low
luminosities points to a solution in which gas joins a jet or outflow  
long before reaching the innermost orbits.
High-resolution radio images of  M87
indeed reveal a very wide base for the jet, of order $100R_S$ or more
(Junor et al. 1999; Ly et al. 2007), suggestive of this picture. 

From a theoretical perspective, a recent model that may produce the 
essential features required by these data is the coronal outflow dominated
accretion disc by Merloni \& Fabian (2002).
In this model, the magnetic stresses inside 
a geometrically thin, optically thick, disc generate
an unbound magnetic corona. 
The corona comptonizes the thermal emission from the
disc, producing the X-ray emission, and serves as a base for
launching a vertical outflow or jet. At progressively lower accretion rates
on to the BH, the ratio of the thermal to comptonized X-ray
emission decreases, and the fraction of the gravitational potential 
energy channelled into 
kinetic energy of the outflow increases. Figure~\ref{alphaoxfig} 
shows, alongside the data, some of these models (kindly provided by
A. Merloni) in the $\alpha_{\rm
  ox}-\nu L\nu(2500~{\rm \AA})$ plane.
Each
curve in the plot is labelled by its BH mass, $\log (M_{\rm
  BH}/M_{\odot})$. The accretion rate along each curve
decreases from right to left, producing a bolometric luminosity of
$10^{-2}L_E$ on the right, and going to $10^{-3.5}L_E$ on the
left. Models are shown for two values of the numerical factor $K$,  
which contains the unknown quantities in the model -- the viscosity, 
and the efficiency of buoyant transport of magnetic structure in the 
vertical direction inside the disc. $K$ is thus 
related to the maximal fraction of power released in the corona.
The top three curves are for $K=0.95$ and the bottom three are for
$K=0.775$. Merloni \& Fabian (2002) note that, at accretion rates
below those corresponding to $L=10^{-3.5}L_E$, the spectral shape
remains unchanged, and the total luminosity simply goes down.
From Fig.~\ref{alphaoxfig}, it appears that large values of $K$
(corresponding to high viscosity and high vertical speed) can lead to 
$\alpha_{\rm ox}$ values comparable to those observed for the LINERs
in the present sample. Furthermore, as the accretion rate decreases,
the outflow becomes energetically more dominant. The radiation from 
this outflow/jet could then lead to the observed increase in $R_{\rm UV}$  
at low luminosities. 

Like many previous accretion models, the Merloni \& Fabian (2002) 
model begins with a dynamical scenario, makes some assumptions,
 and includes free parameters that can be
adjusted to achieve agreement with
observations. An alternative approach has been taken recently by
Loeb \& Waxman (2007), who analyse the millimetre-band emission from
Sag A*. Using measurements of size vs. wavelength of Sag A*, they
conclude that, irrespective of the dynamics, geometry, and physical details 
of the gas flow, a net inflow toward the BH is likely possible only in the 
inner tens of Schwarzschild radii, whereas a net outflow is suggested at
larger radii. Sag A* has an even lower Eddington ratio than the LINERs 
considered in this work, and it is of course unknown what its optical spectrum
would be (LINER or otherwise), 
were we to view it from outside the obscuring disc of the Galaxy.
None the less, the same theme recurs in the Loeb \& Waxman (2007)
analysis -- a low net accretion rate 
on to the BH, with most of the accreting mass at large radii ending up 
in an outflow. This picture is in contrast to that invoked in many
RIAF models, where the accretion energy of the mass inflow is advected
into the BH. 

To summarize, I have compiled recent radio, UV, and X-ray data for a
sample of 13 unobscured low-luminosity LINER AGNs. I have shown that their
interband luminosity ratios are not dramatically different from those
of higher luminosity AGNs. Specifically, in terms of their UV/X
ratios, there is only a slightly enhanced prominence of the X-ray
emission compared to intermediate-luminosity AGNs. There is thus no obvious
indication for the disappearance of the big blue bump at low
luminosities, suggesting the persistence of thin accretion discs
in the low-accretion-rate regime. In terms of radio/UV luminosity
ratios, the LINERs span a range of 4 orders of magnitude, with 
most of them residing at the lower end of the distribution, 
with $\log R_{\rm uv}\la 2$. In this sense, these low-luminosity
AGNs again are part of a continuous sequence with higher luminosity
objects. Since at least some, if not all, of the radio emission in
AGNs is known to come from jets, this suggests a picture in which,
at decreasing accretion rates, a progressively larger fraction of
the inflowing mass is channeled into an outflow, and a smaller
fraction into the persistent thin accretion disk. Analogous results
have been obtained recently for some Galactic stellar-mass BHs. 

While I have made an effort to compile the best available data, the
measurements analyzed here are still crude. In particular, the fact that 
the measurements in different bands are often separated by years, coupled
with the large fluctuations in flux that are common in low-luminosity
AGNs, means that individual luminosity ratios could be off by an
order of magnitude. Future improvements could be achieved by means of
contemporaneous X-ray and UV data, which could possibly be
obtained using the multi-band capabilities on {\it Newton-XMM} or
{\it SWIFT} (for sources with an X-ray flux $\ga 10^{-13}\ergcms$), 
and could be complemented by simultaneous ground-based
radio data. Repeated observations of individual objects over few-year
time-scales could reveal changes in SED as a function
of changing luminosity. Such monitoring would also reduce the
uncertainty regarding the non-stellar contribution to the UV, another
major source of error in the present work. Detection of
nuclear variability also in the optical (with HST) and in the IR
(with {\it Spitzer}) would permit measuring reliably the AGN component
in those bands as well, significantly sharpening our view of the SED.
 Finally, much larger 
samples of unobscured low-luminosity AGNs, analyzed in similar ways, could 
clarify the picture considerably.
 A larger, UV-selected, sample of such objects could
be assembled by means of UV imaging (e.g., with {\it GALEX} or with
HST) and
subsequent optical spectroscopic classification to identify the LINERs. 
UV imaging need not necessarily be from space -- the UV nuclei of the
current sample are prominent in the F330W band, so such objects could
potentially be indentified by ground-based 
observations near the atmospheric UV cutoff. Alternatively,
one could begin with the large, optically selected, 
SDSS LINER sample of Kewley et al. (2006), and follow it up with
sensitive multiwavelength observations.

\section*{Acknowledgments}

I thank Ari Laor, Brent Groves, Marco Chiaberge, Luis Ho, and the
 anonymous referee,
 for very useful suggestions and discussions.
Andrea Merloni is thanked for making his models available in digital form.
This research has made use of the NASA/IPAC Extragalactic Database
(NED) which is operated by the JPL, 
Caltech, under contract with NASA.


\end{document}